\documentclass[epj, nopacs]{svjour}
\onecolumn

\makeatletter
\let\cl@chapter\undefined
\makeatletter

\usepackage{lmodern}
\usepackage[normalem]{ulem}

\usepackage{graphicx}
\usepackage{dcolumn}
\usepackage{bm}

\usepackage{float}
\usepackage{amsmath}
\usepackage[bookmarksdepth=5, hidelinks]{hyperref}
\hypersetup{pageanchor=false}
\usepackage{amssymb}
\usepackage{cleveref}
\usepackage{graphicx}
\usepackage{breqn}
\usepackage{mathtools}
\usepackage{listings}
\usepackage{comment}
\usepackage{multicol}

\usepackage{xcolor}
\usepackage[numbers,sort&compress,square]{natbib}

\newcommand{\infinity}{\infty}
\crefname{figure}{Fig.}{Figs.}
\Crefname{figure}{Figure}{Figures}
\crefname{equation}{Eq.}{Eqs.}
\Crefname{equation}{Equation}{Equations}
\crefname{section}{Sec.}{Secs.}
\Crefname{section}{Section}{Sections}
\creflabelformat{equation}{#2\textup{#1}#3}

\begin{document}

\title{\bfseries\boldmath Inconsistencies in, and short pathlength correction to, \\$R_{AA}(p_T)$ in $\mathrm{A}+\mathrm{A}$ and $\mathrm{p} + \mathrm{A}$ collisions}

\author{Coleridge Faraday\thanks{frdcol002@myuct.ac.za (corresponding)}\and Antonia Grindrod\thanks{agrindrod@student.ethz.ch}\and W.\ A.\ Horowitz\thanks{wa.horowitz@uct.ac.za}}
\institute{Department of Physics\char`,{} University of Cape Town\char`,{} Private Bag X3\char`,{} Rondebosch 7701\char`,{} South Africa}

\abstract{
We present the first leading hadron suppression predictions in $\mathrm{Pb}+\mathrm{Pb}$ and $\mathrm{p}+\mathrm{Pb}$ collisions from a convolved radiative and collisional energy loss model in which partons propagate through a realistic background and in which the radiative energy loss receives a short pathlength correction. We find that the short pathlength correction is small for $D$ and $B$ meson $R_{AA}(p_T)$ in both $\mathrm{Pb}+\mathrm{Pb}$ and $\mathrm{p}+\mathrm{Pb}$ collisions.  However the short pathlength correction leads to a surprisingly large reduction in suppression for $\pi$ mesons in $\mathrm{p}+\mathrm{Pb}$ and even $\mathrm{Pb}+\mathrm{Pb}$ collisions.  We systematically check the consistency of the assumptions used in the radiative energy loss derivation—such as collinearity, softness, and large formation time—with the final numerical model. While collinearity and softness are self-consistently satisfied in the final numerics, we find that the large formation time approximation breaks down at modest to high momenta $p_T \gtrsim 30~\mathrm{GeV}$. We find that both the size of the small pathlength correction to $R_{AA}(p_T)$ and the $p_T$ at which the large formation time assumption breaks down are acutely sensitive to the chosen distribution of scattering centers in the plasma.}

\date{Received: 29 May 2023 / Accepted: 6 November 2023}

\authorrunning{Faraday et al.}
\titlerunning{Inconsistencies in, and short pathlength correction to, $R_{AA}(p_T)$ in $\mathrm{A}+\mathrm{A}$ and $\mathrm{p} + \mathrm{A}$ collisions}

\maketitle

\begin{multicols}{2}
\section{\label{sec:introduction} Introduction}
The modification of the spectrum of high transverse momentum (high-$p_T$) particles is one of the key observables used to understand the non-trivial, emergent, many-body dynamics of quantum chromodynamics (QCD) in high-energy collisions \cite{Gyulassy:2004zy,Wiedemann:2009sh,Majumder:2010qh,Busza:2018rrf}.  One of the most important findings of the Relativistic Heavy Ion Collider (RHIC) was a roughly factor of five suppression of leading light hadrons with $p_T\gtrsim 5$ GeV/c in central $\mathrm{Au}+\mathrm{Au}$ collisions \cite{PHENIX:2001hpc,STAR:2003pjh}.  This suppression, equal for pions and eta mesons \cite{PHENIX:2006ujp}, along with null controls of qualitatively no suppression of the weakly-coupled photons in $\mathrm{Au}+\mathrm{Au}$ collisions \cite{PHENIX:2005yls} as well as of leading hadrons in $\mathrm{d}+\mathrm{Au}$ collisions \cite{STAR:2003pjh,PHENIX:2006mhb}, clearly demonstrated that the suppression of leading hadrons in central collisions is due to final state energy loss of the high-$p_T$ partons interacting with the quark-gluon plasma (QGP) generated in the heavy ion collisions (HIC).  Models of leading hadron suppression based on final state energy loss derivations using perturbative QCD (pQCD) methods qualitatively describe a wealth of these high-$p_T$ RHIC data \cite{Dainese:2004te,Schenke:2009gb,Horowitz:2012cf}.

\sloppy One of the other major findings of RHIC was the near perfect fluidity of the strongly-coupled low momentum modes of the QGP formed in semi-central nucleus-nucleus collisions as inferred by the remarkable agreement of sophisticated, relativistic, viscous hydrodynamics models with the spectra of measured hadrons with $p_T\lesssim2$ GeV \cite{Romatschke:2007mq,Song:2007ux,Schenke:2010rr}.

The data from the Large Hadron Collider (LHC) has been no less impressive.  Of extraordinary importance have been the signs that the non-trivial, emergent, many-body QCD behavior associated with QGP formation in central and semi-central heavy ion collisions at RHIC and LHC are \emph{also} observed in small collision systems such as $\mathrm{p}+\mathrm{p}$ and $\mathrm{p}+\mathrm{A}$ for large final measured multiplicity.  For example, strangeness enhancement \cite{ALICE:2013wgn,ALICE:2015mpp} and quarkonium suppression \cite{ALICE:2016sdt} appear to depend only on multiplicity but not collision system.  And the same sophisticated, relativistic, viscous hydrodynamics models \cite{Weller:2017tsr} also describe the spectra of measured hadrons in high-multiplicity $\mathrm{p}+\mathrm{p}$ and $\mathrm{p}+\mathrm{A}$ collisions \cite{CMS:2015yux,ATLAS:2015hzw}.  

One may thus conclude that small droplets of QGP form even in these smallest of collision systems at the LHC.  If QGP is formed in high-multiplicity collisions of small systems, then high-$p_T$ partons should suffer some final state energy loss, as has been observed in large collision systems at RHIC and LHC.  (Models already demonstrate the importance of final state energy loss in forward hadron production in cold nuclear matter \cite{Arleo:2020eia,Arleo:2021bpv}.)  Experimentally, there are tantalizing signs of the non-trivial modification of high-$p_T$ particles in small collision systems \cite{ATLAS:2014cpa,PHENIX:2015fgy,ALICE:2017svf}.  However, there are likely non-trivial correlations between the multiplicity in small collision systems and the presence of high-$p_T$, high-multiplicity jets.  For example, these correlations likely impact the initial spectrum of high-$p_T$ partons \cite{ALICE:2018pal} that enters the numerator of the nuclear modification factor, while the minimum bias spectrum in the denominator is unchanged; thus one should be cautious in interpreting a standard $R_{AB}$ measurement.  

\sloppy From the experimental side, it is likely very interesting to consider the ratio of the spectrum of a strongly-interacting particle with a weakly-interacting particle, each from the same multiplicity class.

From the theoretical side, we may potentially make progress by considering the small collision system predictions of the energy loss models used to describe so well qualitatively the large collision system high-$p_T$ particle suppression.  One obvious challenge for directly comparing these energy loss models to small collision system data is the assumption made in the energy loss derivations that the high-$p_T$ particles travel a large pathlength in the QGP medium. For example, energy loss models built on BDMPS-Z-type energy loss \cite{Baier:1996kr, Baier:1996sk, Baier:1996vi, Baier:1998kq,Zakharov:1996fv,Zakharov:1997uu} utilize the central limit theorem \cite{Armesto:2011ht}, and so assume a very large number of collisions occur between the high-$p_T$ probe and the QGP medium.  Even in large collision systems of characteristic size $\sim5$ fm, since the mean free path for these high-$p_T$ particles given by weakly-coupled pQCD is $\sim1\text{--}2$ fm \cite{Armesto:2011ht}, the application of the central limit theorem is dubious.  

\sloppy Even for the thin plasma approach of DGLV that naively seems the best suited for modelling the radiative energy loss processes in systems of phenomenologically relevant size, there is an explicit assumption that the partonic pathlength in the QGP medium, $L$, is large compared to the natural scale set by the Debye screening mass $\mu$, $L\gg1/\mu$.  In the original derivation of the induced gluon radiation spectrum, contributions from the Gyulassy-Wang potential \cite{Gyulassy:1993hr} were dropped as they are $\mathcal O(e^{-L\mu})$.  For $\mu\sim gT\sim 0.5$ GeV \cite{Kapusta:2006pm}, the characteristic size $L\sim 1$ fm in high-multiplicity $\mathrm{p}+\mathrm{p}$ and $\mathrm{p}+\mathrm{A}$ collisions is not particularly large compared to $1/\mu\sim0.4$ fm.  Thus to create an energy loss model to compare to data in these small systems, one needed the small pathlength corrections to the DGLV opacity expansion.

This small pathlength correction to the first order in opacity DGLV radiative energy loss were derived for the first time in \cite{Kolbe:2015rvk}.  
For later noting, the derivation in \cite{Kolbe:2015rvk} benefited significantly from a simplification due to the large formation time assumption, $\tau_{\text{form}}\gg1/\mu$, an assumption made also in the original DGLV derivations.

The small pathlength correction to the usual DGLV energy loss contained four surprises: due to the LPM effect (interference between the induced radiation and the usual vacuum emissions due to the hard initial scattering), the small pathlength correction \emph{reduces} the energy lost; the reduction in energy loss is seen in \emph{all} pathlengths (although the relative importance of the correction decreases with pathlength, as expected); the correction \emph{grows linearly} with partonic energy (as opposed to the logarithmic growth of the usual DGLV energy loss); and the correction breaks color triviality, with the correction for gluons $\sim10$ times the size of the correction for quarks.

Having derived the correction to the radiative energy loss due to small pathlengths, it is of considerable interest to determine quantitatively the importance of the correction in phenomenological models of high-$p_T$ particle suppression.  It is the goal of this manuscript to provide just such predictions.  We are particularly interested in seeing the quantitative importance of the reduction in energy loss as a function of energy and of collision system: the reduction in energy loss from the short pathlength correction might provide a natural explanation for the surprisingly fast rise in the nuclear modification factor with $p_T$ for leading hadrons at LHC \cite{Horowitz:2011gd} and for the enhancement of the nuclear modification factor above unity seen in $\mathrm{p}+\mathrm{A}$ collisions \cite{Balek:2017man,ALICE:2018lyv}.

What we will see is that for light flavor final states at larger energies, $p_T\gtrsim30$ GeV/c, the small system ``correction'' becomes of order of the energy loss itself, which leads us to consider systematically the extent to which energy loss model energy losses are consistent with the various approximations used in the opacity expansion derivation.  We will consider in detail the approximations used in the opacity expansion derivation of radiative energy loss \cite{Vitev:2002pf,Djordjevic:2003zk,Kolbe:2015rvk}.  We will find that for the radiated gluons that dominate the energy weighted single inclusive emission distribution, the large formation time assumption is violated for $E\gtrsim30$ GeV/c in large collision systems and for $E\gtrsim10$ GeV/c in small systems, where $E$ is the energy of the radiating parton; which implies the need for yet another derivation of radiative energy loss in the thin plasma limit but with the large formation time assumption relaxed. We also see that the usual WHDG treatment of the average elastic energy loss, with fluctuations given by the fluctuation-dissipation theorem, is not appropriate for small collision systems where the number of scatterings is not large; thus in future work one must implement an elastic energy loss appropriate for small and large collision systems.

\section{\label{sec:energy_loss} Energy loss framework}

We wish to make contact with known energy loss model results.  We will in particular attempt to, in as reasonable way as possible, make an apples-to-apples comparison of the Wicks-Horowitz-Djordjevic-Gyulassy (WHDG) convolved radiative and collisional energy loss model \cite{Wicks:2005gt}, which has seen such success in describing a breadth of leading hadron suppression data, with an energy loss model with the same elastic energy loss but with a radiative energy loss that includes the short pathlength correction as derived in \cite{Kolbe:2015rvk}.  Let us briefly review the radiative and collisional energy loss setups.

\subsection{Radiative energy loss}
\label{sub:radiative_energy_loss}

The Djordjevic-Gyulassy-Levai-Vitev (DGLV)
opacity expansion \cite{Djordjevic:2003zk, Gyulassy:1999zd}
gives the inclusive differential distribution of radiated gluons from a high-$p_T$ parent parton moving through a smooth brick of QGP. The expansion is in the expected number of scatterings or the \textit{opacity} $L / \lambda_g$, where $L$ is the length of the QGP brick and $\lambda_g$ is the mean free path of a gluon in the QGP.

The 4-momenta of the radiated gluon, the final hard parton, and the exchanged Debye medium quasiparticle are given respectively in lightfront coordinates (using the same conventions as in \cite{Kolbe:2015rvk}) by
\begin{subequations}
\begin{gather}
 k=\left[x P^{+}, \frac{m_g^2+\mathbf{k}^2}{x P^{+}}, \mathbf{k}\right] \\
 p=\left[(1-x) P^{+}, \frac{M^2+\mathbf{k}^2}{(1-x) P^{+}}, \mathbf{q}-\mathbf{k}\right] \\
 q=\left[q^{+}, q^{-}, \mathbf{q}\right],
\label{eqn:four_momenta}
\end{gather}
\end{subequations}
where $M$ is the mass of the hard parton, $m_g$ is the gluon mass, $P^+$ is the initial hard parton momentum in the $+$ direction, and $x$ is the radiated momentum fraction.

The DGLV approach makes a number of assumptions related to the physical setup of the problem:
\begin{itemize}
    \item The large pathlength assumption, that $L \gg \mu^{-1}$.
    \item The well separated scattering centers assumption, that $\lambda_g \gg \mu^{-1}$. %
    \item The eikonal assumption, that $P^+ = E^+ \simeq 2E$ is the largest scale in the problem.
    \item The soft radiation assumption, that $x\ll1$.
    \item The collinear radiation assumption, that $k^+ \gg k^-$.
    \item The large formation time assumption, that $\mathbf{k}^2 /x E^+ \ll \mu$ and $(\mathbf{k}-\mathbf{q}_1)^2/x E^+ \ll \sqrt{\mu^2+\mathbf{q}_1^2}$.
\end{itemize}
The DGLV single inclusive gluon radiation spectrum is then \cite{Gyulassy:2000er,Djordjevic:2003zk}
\end{multicols}
\begin{gather}
  \frac{\mathrm{d} N^g_{\text{DGLV}}}{\mathrm{d} x}=  \frac{C_R \alpha_s L}{\pi \lambda_g} \frac{1}{x} \int \frac{\mathrm{d}^2 \mathbf{q}_1}{\pi} \frac{\mu^2}{\left(\mu^2+\mathbf{q}_1^2\right)^2} \int \frac{\mathrm{d}^2 \mathbf{k}}{\pi} \int \mathrm{d} \Delta z \, \bar{\rho}(\Delta z) \nonumber\\
 \times -\frac{2\left\{1-\cos \left[\left(\omega_1+\tilde{\omega}_m\right) \Delta z\right]\right\}}{\left(\mathbf{k}-\mathbf{q}_1\right)^2+m_g^2+x^2 M^2}\left[\frac{\left(\mathbf{k}-\mathbf{q}_1\right) \cdot \mathbf{k}}{\mathbf{k}^2+m_g^2+x^2 M^2}-\frac{\left(\mathbf{k}-\mathbf{q}_1\right)^2}{\left(\mathbf{k}-\mathbf{q}_1\right)^2+m_g^2+x^2 M^2}\right].
 \label{eqn:DGLV_dndx}
\end{gather}
\begin{multicols}{2}
In \cref{eqn:DGLV_dndx} we have made use of the shorthand $\omega \equiv x E^+ / 2,~\omega_0 \equiv \mathbf{k}^2 / 2 \omega,~\omega_i \equiv (\mathbf{k} - \mathbf{q}_i)^2 / 2 \omega$, $\mu_i \equiv \sqrt{\mu^2 + \mathbf{q}_i^2}$, and $\tilde{\omega}_m \equiv (m_g^2 + M^2 x^2) / 2 \omega$ following \cite{Djordjevic:2003zk, Kolbe:2015rvk}. Additionally $\mathbf{q}_i$ is the transverse momentum of the $i^{\mathrm{th}}$ gluon exchanged with the medium; $\mathbf{k}$ is the transverse momentum of the radiated gluon; $\Delta z$ is the distance between production of the hard parton, and scattering; $C_R$ ($C_A$) is the quadratic Casimir of the hard parton (adjoint) representation ($C_F = 4 / 3$ [quarks], and $C_A = 3$ [gluons]); and $\alpha_s$ is the strong coupling.

The quantity $\bar{\rho}(\Delta z)$ is the \emph{distribution of scattering centers} in $\Delta z$ and is defined in terms of the density of scattering centers $\rho(\Delta z)$ in a static brick,
\begin{equation}
  \rho(\Delta z) = \frac{N}{A_{\perp}} \bar{\rho}(\Delta z),
  \label{eqn:density_scattering_centers}
\end{equation}
where $\Delta z$ is in the direction of propagation, $N$ is the number of scattering centers, $A_{\perp}$ is the perpendicular area of the brick, and $\int \mathrm{d}z \; \bar{\rho}(\Delta z) = 1$. The analysis of realistic collision geometries adds complexity to the scenario, as detailed in \cref{sec:geometry}.

\subsection{Short pathlength correction to DGLV radiative energy loss}
\label{sub:radiative_energy_loss_correction}
The derivation of the modification to the radiative energy loss in the DGLV \cite{Vitev:2002pf,Djordjevic:2003zk} opacity expansion approach with the relaxation of the large pathlength assumption $L \gg \mu^{-1}$ was considered in \cite{Kolbe:2015rvk,Kolbe:2015suq}.  In the derivation of the short pathlength correction, all assumptions and approximations made in the original GLV and DGLV derivations were kept, except that the short pathlength approximation $L\gg\mu^{-1}$ was relaxed.  The single inclusive radiative gluon distribution, including both the original DGLV contribution as well as the short pathlength correction, is
\end{multicols}
\begin{gather}
    \frac{\mathrm{d} N^g_{\text{DGLV+corr}}}{\mathrm{d} x}=  \frac{C_R \alpha_s L}{\pi \lambda_g} \frac{1}{x} \int \frac{\mathrm{d}^2 \mathbf{q}_1}{\pi} \frac{\mu^2}{\left(\mu^2+\mathbf{q}_1^2\right)^2} \int \frac{\mathrm{d}^2 \mathbf{k}}{\pi} \int \mathrm{d} \Delta z \, \bar{\rho}(\Delta z) \nonumber\\
   \times\left[-\frac{2\left\{1-\cos \left[\left(\omega_1+\tilde{\omega}_m\right) \Delta z\right]\right\}}{\left(\mathbf{k}-\mathbf{q}_1\right)^2+m_g^2+x^2 M^2}\left[\frac{\left(\mathbf{k}-\mathbf{q}_1\right) \cdot \mathbf{k}}{\mathbf{k}^2+m_g^2+x^2 M^2}-\frac{\left(\mathbf{k}-\mathbf{q}_1\right)^2}{\left(\mathbf{k}-\mathbf{q}_1\right)^2+m_g^2+x^2 M^2}\right] \right. \nonumber\\
   +\frac{1}{2} e^{-\mu_1 \Delta z}\left(\left(\frac{\mathbf{k}}{\mathbf{k}^2+m_g^2+x^2 M^2}\right)^2\left(1-\frac{2 C_R}{C_A}\right)\left\{1-\cos \left[\left(\omega_0+\tilde{\omega}_m\right) \Delta z\right]\right\}\right. \nonumber\\
   \left.\left.+\frac{\mathbf{k} \cdot\left(\mathbf{k}-\mathbf{q}_1\right)}{\left(\mathbf{k}^2+m_g^2+x^2 M^2\right)\left(\left(\mathbf{k}-\mathbf{q}_1\right)^2+m_g^2+x^2 M^2\right)}\left\{\cos \left[\left(\omega_0+\tilde{\omega}_m\right) \Delta z\right]-\cos \left[\left(\omega_0-\omega_1\right) \Delta z\right]\right\}\right)\right],
   \label{eqn:full_dndx}
\end{gather}
\begin{multicols}{2}
\noindent where the first two lines of the above equation are the original DGLV result \cite{Gyulassy:2000er,Djordjevic:2003zk}, \cref{eqn:DGLV_dndx}, while the last two lines are the short pathlength correction.
We emphasize that contributions from all diagrams which are not suppressed under the relevant assumptions are included.  Of particular importance is the large formation time assumption, which allows one to systematically neglect a significant number of diagrams in both the original DGLV derivation \cite{Vitev:2002pf,Djordjevic:2003zk} and in the short pathlength correction \cite{Kolbe:2015rvk,Kolbe:2015suq}.

Since $\mathrm{d} N/\mathrm{d} x$ includes an integration over all $\Delta z$, the correction is present for the energy loss of a parton going through plasma of \emph{any} length; however, the relative contribution of the correction term does go to zero as the pathlength goes to infinity.

The finite pathlength correction originates from not neglecting the $q^z = i\mu_1$ pole in the Gyulassy-Wang potential, as was originally done \cite{Gyulassy:2000er,Djordjevic:2003zk}, which leads to the overall $\exp(-\mu_1\Delta z)$ scaling of the correction term in \cref{eqn:full_dndx} \cite{Kolbe:2015rvk,Kolbe:2015suq}.  

There is a significant literature of energy loss derivations and corrections to earlier energy loss derivations.  Even though the focus of this work is the numerical implementation of \cref{eqn:full_dndx} and the examination of its underlying assumptions, it is worth taking some time to contextualize the short pathlength correction in \cref{eqn:full_dndx} within the literature.  In particular, there is currently no other derivation of short pathlength corrections to any energy loss formalism in the literature.  

The original BDMPS \cite{Baier:1996kr, Baier:1996sk, Baier:1996vi, Baier:1998kq} energy loss derivation explicitly neglects the $q^z=i\mu$ pole in the Gyulassy-Wang potential.  In principle, then, one could  derive a short pathlength correction in the original BDMPS-Z formalism analogous to the one derived in \cite{Kolbe:2015rvk,Kolbe:2015suq}.  Subsequent work within the BDMPS-Z formalism \cite{Baier:1996kr, Baier:1996sk, Baier:1996vi, Baier:1998kq,Zakharov:1996fv,Zakharov:1997uu} considered the saddle point approximation of the path integral, that in the limit of a large number of scatterings one could make a simple harmonic oscillator approximation (via the Central Limit Theorem).  This SHO approximation explicitly requires a large opacity $L / \lambda \gg 1$.  For a perturbative calculation, one requires the scattering centers are well-separated, $\lambda\gg1/\mu$, and so a large opacity implies a large pathlength; one therefore cannot determine a short pathlength correction to the SHO approximated BDMPS-Z approach.  %
If one assumes that the system is strongly coupled (see, e.g., \cite{Liu:2006ug,Casalderrey-Solana:2011dxg}) and $\lambda\ll\mu^{-1}$, then all paths are long and there is no short pathlength correction.  

In the Improved Opacity Expansion (IOE) \cite{Mehtar-Tani:2019ygg,Mehtar-Tani:2019tvy}, the starting point is already the $z$-integrated path integral; i.e.\ the IOE starts with the equation of motion for the propagator in transverse position space, and is completely insensitive to questions about the interplay between any longitudinal scales such as the pathlength, mean free path, and distance between scattering centers.  Within the IOE formalism the ``Gyulassy-Wang'' potential is taken to be $V\left(\mathbf{q}^2\right) \sim 1 / \left(\mathbf{q}^2+\mu^2\right)^2$ where, importantly, $\mathbf{q}$ is the transverse momentum exchanged with the scattering center; i.e.\ any potential pole from the $z$ component of the fully three dimensional Gyulassy-Wang potential is already neglected.  Thus the IOE formalism is unable to compute any short pathlength correction to the energy loss. %

Similar to the IOE approach, the finite size-improvement \cite{Caron-Huot:2010qjx} to the AMY formalism \cite{Arnold:2001ba,Arnold:2001ms,Arnold:2002ja} begins with Zakharov's path integral formalism and considers only the transverse momentum transfer $\mathbf{q}$ from the in-medium scattering cross section.  Thus, like in the IOE approach, unless the scattering center cross section is considered in full three dimensions, the finite size-improvement to AMY cannot capture the short pathlength corrections to energy loss.

There is a further extensive literature of work that utilizes only the 2D (rather than 3D) potential, and thus cannot capture the short pathlength corrections found in \cite{Kolbe:2015rvk,Kolbe:2015suq}.  Some of these works include the antenna problem \cite{Mehtar-Tani:2010ebp,Mehtar-Tani:2011hma,Mehtar-Tani:2011vlz,Armesto:2011ir,Mehtar-Tani:2011lic,Mehtar-Tani:2012mfa}, $\hat q$ resummation \cite{Blaizot:2014bha}, jet cascades \cite{Blaizot:2013vha,Iancu:2014kga}, running coupling effects in $\hat q$ \cite{Iancu:2014sha}, the radiative energy loss of neighboring subjets \cite{Mehtar-Tani:2017ypq}, and quark branching beyond the soft gluon limit \cite{Sievert:2018imd}.  

In \cite{Sadofyev:2021ohn} the authors couple the opacity expansion to the collective flow of the quark gluon plasma.  This work explicitly makes the assumption $\mu \Delta z \gg 1$; thus short pathlength corrections to this derivation are possible, but not included in the original derivation.

In \cite{Djordjevic:2007at, Djordjevic:2008iz, Djordjevic:2009cr} the Gyulassy-Wang static scattering center potential in canonical opacity expansion calculations was replaced by HTL propagators communicating between the high-energy parent parton and the in-medium thermal parton.  In the first derivation \cite{Djordjevic:2007at, Djordjevic:2008iz}, the authors work purely in momentum space to compute the interaction rate; since the authors do not Fourier transform into position space, their result is completely insensitive to the particulars of the path.  This work was improved in \cite{Djordjevic:2009cr} where a Fourier transform into position space was done and, as is done in the opacity expansion approach, the phases were kept.  However the limit $L\to\infinity$ is explicitly taken.  In principle one could derive a short pathlength correction to this derivation by relaxing the assumption of $L\to\infinity$.

The Higher Twist (HT) approach \cite{Guo:2000nz,Wang:2001ifa} in general only keeps the most length-enhanced contributions \cite{Majumder:2007hx}.  In principle, one may include less enhanced contributions from the assumed factorized nuclear expectation values of the various four point functions.

Gradient jet tomography \cite{He:2020iow,Fu:2022idl} couples a high momentum parton to an asymmetric medium.  The practical implementation of this procedure utilizes the dipole approximation in the path integral approach where the entire nuclear medium is treated as a sheet \cite{Casalderrey-Solana:2007knd}, and so any sensitivity to longitudinal physics is lost.

\subsection{Numerical implementation of the radiative energy loss}

For all numerical calculations we neglect the running of the strong coupling constant, and use $\alpha_s = 0.3$, consistent with \cite{Kolbe:2015rvk, Wicks:2005gt, Djordjevic:2003zk}. Additionally we use charm and bottom quark masses of $m_c = 1.2~\mathrm{GeV}/c^2$, and $m_b = 4.75~\mathrm{GeV}/c^2$, respectively. The effective light quark and gluon masses are set to the asymptotic one-loop medium induced thermal masses, of $m_{\text{light}} = \mu / 2$ and $m_g = \mu / \sqrt{2}$ respectively \cite{Djordjevic:2003be}. The upper bounds on the $|\mathbf{k}|$ and $|\mathbf{q}|$ integrals are given by $k_{\text{max}}=2x (1-x) E$ and $q_{\text{max}}=\sqrt{3 E \mu}$, following \cite{Wicks:2005gt}. This choice of $k_{ \text{max}}$ guarantees that the momentum of the radiated gluon; and the initial and final momenta of the parent parton are all collinear.

In order to maintain consistency with the WHDG model, we assume the distribution of scattering centers is exponential: $\rho_{\text{exp.}} (z) \equiv \frac{2}{L} \exp[ - 2 z / L]$. The exponential distribution serves to make the integral in $\Delta z$ analytically simple. The physical motivation for this is that an exponentially decaying distribution of scattering centers captures the decreasing density of the QGP due to Bjorken expansion \cite{Bjorken:1982qr}. It is likely that using an exponential rather than, say, a unit step or power law decay distribution is an overestimate of the effect of the expansion, since Bjorken expansion obeys a power law decay along the incident partons path, not exponential. 
It turns out that, for the uncorrected DGLV result, the distribution of scattering centers $\bar{\rho}(\Delta z)$ affects the characteristic shape of $\mathrm{d} N^g / \mathrm{d} x$; however we will see that the radiative energy loss is largely insensitive to $\bar{\rho}(\Delta z)$ \cite{Armesto:2011ht}. Once the short pathlength correction is included, however, the energy loss becomes far more sensitive to the distribution of scattering centers; particularly at small $\Delta z$ \cite{Kolbe:2015rvk}.

The integral in \cref{eqn:full_dndx} can be dramatically simplified if one assumes that the kinematic bound on $q$, $q_{\text{max}} = \sqrt{3 E \mu}$, can be taken to infinity. Assuming that $q_{\text{max}}\rightarrow\infinity$ allows one to perform the angular and $\mathbf{k}$ integrals analytically, after which one may return the kinematic limit $q_{\text{max}} = \sqrt{3 E \mu}$. This procedure was done for all previous WHDG predictions \cite{Wicks:2005gt, Horowitz:2011gd, Andronic:2015wma} in order to make the numerics simpler; however we performed the full numerical calculation without using this approximation. A full description of this kinematic approximation is provided in \cite{Djordjevic:2003zk}.

\begin{figure}[H]
    \centering
    \includegraphics[width=\linewidth]{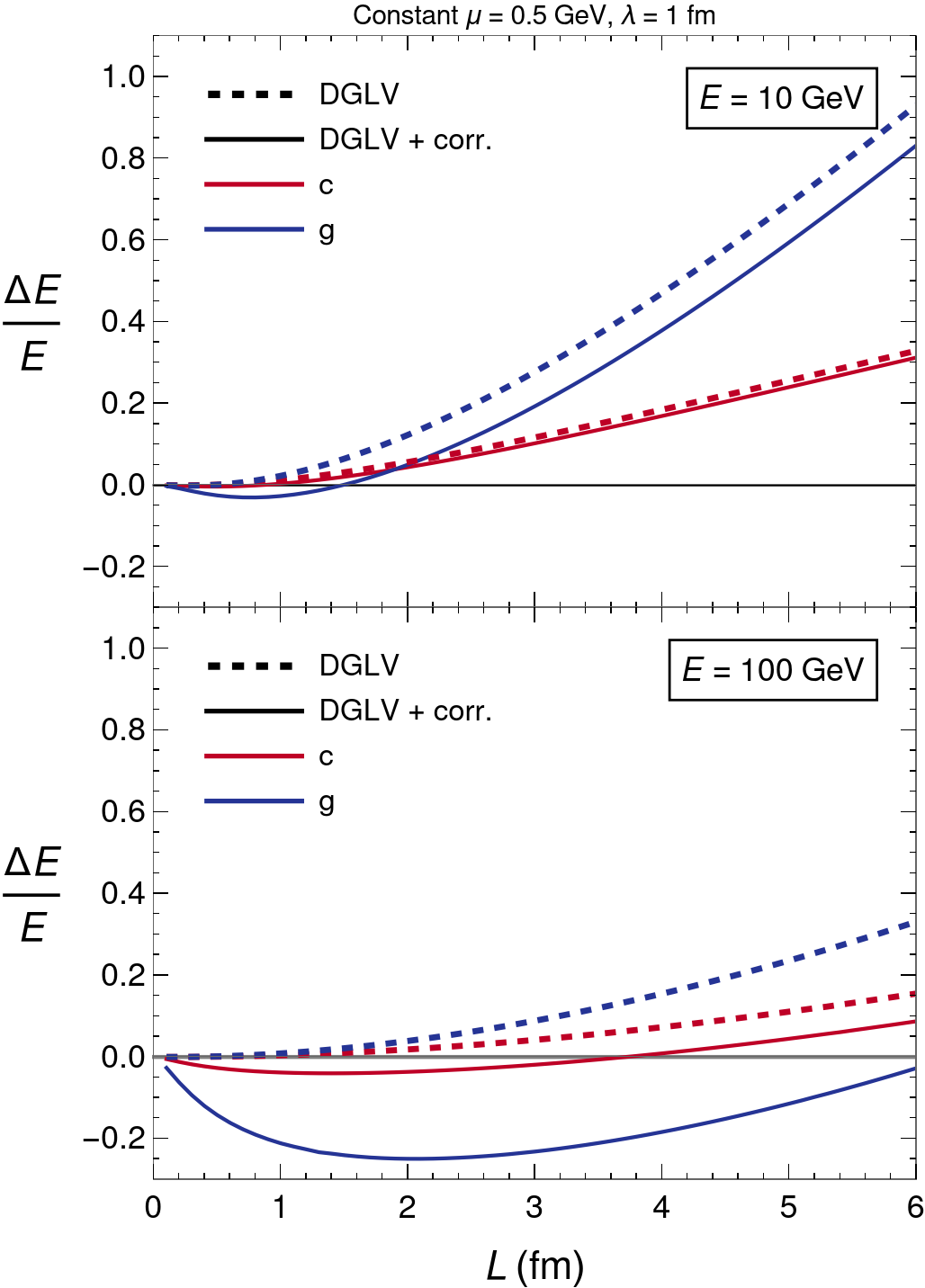}
    \caption{The induced radiative fractional energy loss $\Delta E / E$ is plotted as a function of energy $E$ for charm quarks (c), and gluons (g); and at $E=10 \; \mathrm{GeV}$ (top panel), and $E = 100\; \mathrm{GeV}$ (bottom panel). Note that $\Delta E / E < 0$ is energy gain relative to the vacuum. Calculations were done with constant $\mu = 0.5 ~\mathrm{GeV}$ and $\lambda_g = 1~\mathrm{fm}$. %
    }
    \label{fig:deltaEOverEShortvsDGLV}
\end{figure}

To gain some familiarity with the effect of the short pathlength correction, we calculate the fractional radiative energy loss numerically according to \cref{eqn:full_dndx}. The mean radiative energy loss is calculated from $\mathrm{d} N^g / \mathrm{d} x$ using
\begin{equation}
    \left\langle \frac{\Delta E}{E} \right\rangle = \int \mathrm{d} x \; x \frac{\mathrm{d} N^g}{ \mathrm{d} x}.
    \label{eqn:fractional_energy_loss}
\end{equation}
The results of evaluating \cref{eqn:fractional_energy_loss} as a function of the medium length $L$ are shown in \cref{fig:deltaEOverEShortvsDGLV}. The four surprises mentioned in the Introduction are in evidence here: the short pathlength correction \emph{reduces} the energy loss; the short pathlength correction does not disappear for $L \gg \mu^{-1}$, since all possible distances of scattering $\Delta z$ are integrated over in \cref{eqn:fractional_energy_loss}; the short pathlength correction grows linearly with energy; and the effect of the \emph{reduction} in vacuum radiation is particularly strong for gluons because of the breaking of color triviality. 

\begin{figure}[H]
    \centering
    \includegraphics[width=\linewidth]{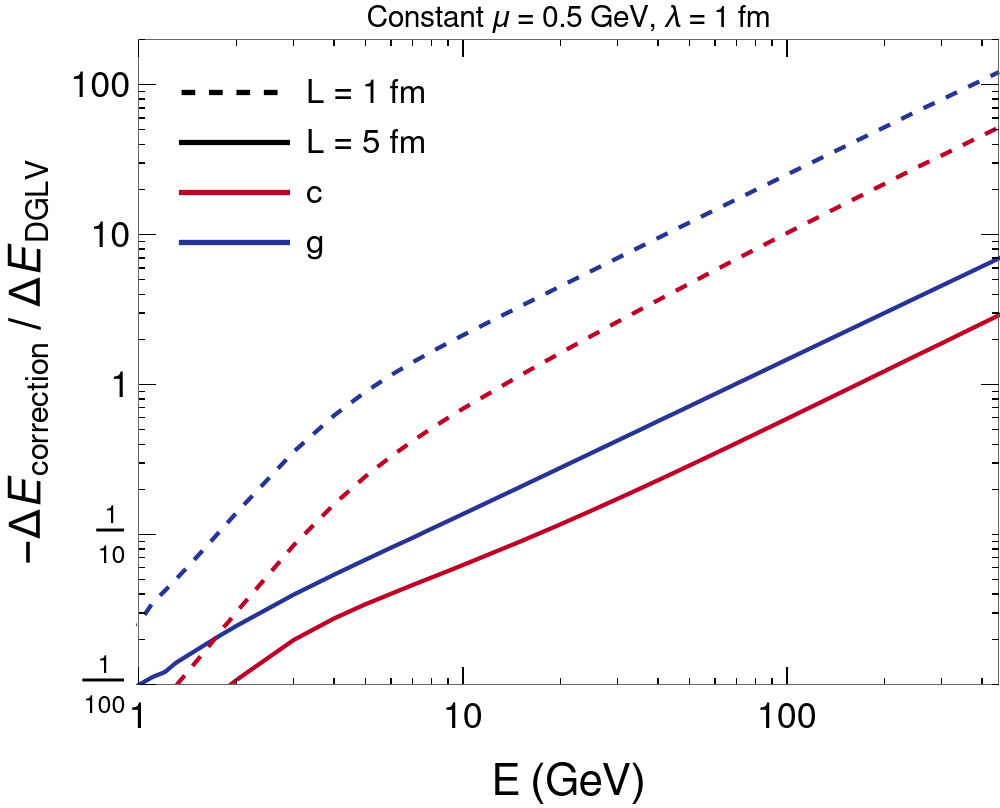}
    \caption{The ratio of the magnitude of the correction to the DGLV radiative energy loss $-\Delta E \text{corrected}$ and the uncorrected DGLV radiative energy loss $\Delta E_{\text{DGLV}}$ is plotted as a function of incident energy $E$. This ratio is plotted for charm quarks (c), and gluons (g); and at $L = 1~\mathrm{fm}$ (dashed) and $L= 5~\mathrm{fm}$ (solid). Calculations were done with constant $\mu = 0.5 ~\mathrm{GeV}$ and $\lambda_g = 1~\mathrm{fm}$.}
    \label{fig:deltaE_vs_energy}
\end{figure}

One may also consider the relative size of the short pathlength correction to the radiative energy loss in comparison to the uncorrected radiative energy loss, by calculating the ratio $\Delta E_{\text{corrected}} / \Delta E_{\text{DGLV}}$, shown in \cref{fig:deltaE_vs_energy}.

  To understand the energy $E$, length $L$, and incident Casimir $C_R$ dependence of the relative size of the short pathlength correction to the radiative energy loss, we examine the asymptotic dependence of both the DGLV and short pathlength corrected DGLV fractional radiative energy loss. For asymptotic energies the short pathlength corrected fractional energy loss is given by \cite{Kolbe:2015rvk}
\begin{subequations}
\begin{align}
  \frac{\Delta E_{\text{corrected}}}{E} =& \frac{C_R \alpha_s}{2 \pi} \frac{L}{\lambda_g}\left(-\frac{2 C_R}{C_A}\right) \frac{12}{2+\mu L}\nonumber\\
  &\times \int_0^1 \mathrm{d} x \ln \left(\frac{L k_{\max }}{2+\mu L}\right)
  \label{eqn:asymptotics_correction_xint}\\
  =& \frac{C_R \alpha_s}{2 \pi} \frac{L}{\lambda_g}\left(-\frac{2 C_R}{C_A}\right) \frac{\ln \left(\frac{2 E L}{2+\mu L}\right)}{2+\mu L},
  \label{eqn:asymptotics_correction}
\end{align}
\end{subequations}
which was calculated in this asymptotic analysis for simplicity with $k_{\text{max}}=2x E$ and, also for simplicity, an exponential distribution of scattering centers. The equivalent asymptotic result without the short pathlength correction is given by \cite{Gyulassy:2000er}
\begin{equation}
  \frac{\Delta E_{\text{DGLV}}}{E} = \frac{C_R \alpha_s}{4} \frac{L^2 \mu^2}{\lambda_g} \frac{1}{E}\log \frac{E}{\mu}.
    \label{eqn:asymptotics_DGLV}
\end{equation}
The relative size of the correction is then given by
\begin{align}
  \frac{-\Delta E_{\text{correction}}}{\Delta E_{\text{DGLV}}} &= \frac{4}{\pi} \frac{C_R}{C_A} \frac{E}{\mu^2 L(2 + \mu L)},
  \label{eqn:asymptotic_relative_correction}
\end{align}
keeping only leading terms in $E$. From \cref{eqn:asymptotic_relative_correction}, the relative size of the short pathlength correction: increases linearly in energy, is $C_A / C_F = 9 /4$ times larger for gluons in comparison to quarks, and is $\sim 15$ times larger for a system with $\mu = 0.5$ GeV and $L = 1~\mathrm{fm}$ as opposed to a system with $L = 5~\mathrm{fm}$.  One may see in \cref{fig:deltaE_vs_energy} that the detailed numerics display these three behaviors.

\subsection{Multi-gluon emission}
\label{sub:multi_gluon_emission}
The DGLV energy loss kernel \cref{eqn:full_dndx} gives the inclusive spectrum of emitted gluons.  Thus the expected number of gluons can be greater than 1.  In fact, one sees that for hard partons emerging from the center of a central heavy ion collision, the expected number of emitted gluons is $\sim 3$ \cite{Gyulassy:2001nm}. To take into account multi-gluon emission we assume that the multiple gluon emissions are independent, following \cite{Gyulassy:2001nm}. This assumption of independent emissions allows us to convolve the single inclusive gluon emission kernel given by $\mathrm{d} N^g / \mathrm{d} x$ into a Poisson distribution. Explicitly we can write 
\begin{equation}
  P_{\text{rad}}(\epsilon, E)=\sum_{n=0}^{\infty} P_n(\epsilon, E),
  \label{eqn:poisson}
\end{equation}
where the $P_n$ are found via the convolution
\begin{align}
  P_{n+1}(\epsilon) & =\frac{1}{n+1} \int \mathrm{d} x_n \; \frac{\mathrm{d} N^{g}}{\mathrm{d} x} \; P_n(\epsilon-x_n)
    \label{eqn:pn}
\end{align}
and we have $P_0(\epsilon) \equiv e^{- \langle N^g \rangle} \delta(\epsilon)$. Here, and for the rest of the paper, we define $1-\epsilon$ as the fraction of initial momentum kept by the parton, that the final energy of the parton in terms of the initial energy of the parton is $E_f\equiv(1-\epsilon)E_i$.  The Poissonian form of \cref{eqn:poisson} guarantees the distribution is normalized to one, and the expected number of emitted gluons is $\sum_n\int \mathrm{d} \epsilon \: n \, P_n(\epsilon, E) = \langle N^g \rangle$. The bounds on the $x_n$ integral are $\max(0,\epsilon-1) \leq x_n \leq \min(\epsilon,1)$, which are determined by ensuring that no functions are evaluated outside of their domains. The support of $P_{\text{rad}}(\epsilon)$ ends when the bounds of the $x_n$ integral are equal, i.e.\ $\epsilon \in (0, 2)$ is the region of support. The support of $P_{\text{rad}}(\epsilon)$ past $\epsilon = 1$ is unphysical, and we interpret this as the probability for the parton to lose all of its energy before exiting the plasma. Under this interpretation we put the excess weight $\int_1^2 \mathrm{d} \epsilon \; P_{\text{rad}}(\epsilon)$ into a delta function at $\epsilon=1$.

We note that the random variable $N^g$ should rigorously be thought of as $N^g = (N^g_{\text{vac}} + N^g_{\text{med}}) - N^g_{\text{vac}}$ where $N^g_{\text{med}}$ is the number of radiated gluons occurring due to medium interactions, and $N^g_{\text{vac}}$ is the number of DGLAP vacuum radiation gluons. With this understanding, the independent gluon emission assumption means that $(N^g_{\text{vac}} + N^g_{\text{med}})$ and $N^g_{\text{vac}}$ should each be modeled by a Poisson distribution.  Then $N^g$ is actually given by a Skellam distribution \cite{skellam:1946}, the difference between two Poisson distributions.
In the current model, energy gain relative to the vacuum at some $x^*$ corresponds to $\mathrm{d} N^g (x^*) / \mathrm{d} x < 0$; whereas it should rigorously correspond to $\mathrm{d} N^g (-x^*) / \mathrm{d} x > 0$. Following previous work \cite{Gyulassy:2001nm, Wicks:2005gt, Horowitz:2011gd}, we will simply model $P(\epsilon)$ as a Poisson distribution; the effect of this simplification is not obvious and requires future work.

\subsection{Elastic energy loss}
\label{sub:elastic_energy_loss}

Elastic energy loss is taken into account using the result derived by Braaten and Thoma (BT) \cite{Braaten:1991we}, wherein analytic calculations are done in two asymptotic energy regimes. The elastic energy loss of a quark is calculated in the regions $E \ll M^2 / T$ and $E \gg M^2 / T$, where $M$ is the mass of the incident quark, and $T$ is the temperature of the medium. For $E \ll M^2 / T$ the differential energy loss per unit distance is
\begin{multline}
\frac{\mathrm{d} E}{\mathrm{d} z} = \frac{8 \pi \alpha_s^2 T^2}{3}\left[\frac{1}{v}- \frac{1-v^2}{2 v^2}\log \frac{1+v}{1-v}\right]\\
\times \log \left(2^{\frac{n_f}{6+n_f}} B(v) \frac{E T}{m_g M}\right)\left(1+\frac{n_f}{6}\right),
\label{eqn:elastic_energy_loss_low}
\end{multline}
where $B(v)$ is a smooth function satisfying constraints listed in \cite{Braaten:1991we}, $v$ is the velocity of the hard parton, and $n_f$ is the number of active quark flavors in the plasma (taken to be $n_f = 2$ throughout). For $E \gg M^2/T$ the differential energy loss per unit distance is
\begin{multline}
    \frac{\mathrm{d} E}{\mathrm{d} z} = \frac{8 \pi \alpha_s^2 T^2}{3} \left(1 + \frac{n_f}{6}\right) \\
    \times \log \left(2^{\frac{n_f}{2(6+n_f)}} \, 0.92 \frac{\sqrt{E T}}{m_g}\right).
    \label{eqn:elastic_energy_loss_high}
\end{multline}
The energy loss at arbitrary incident energy is taken to be the connection of these two asymptotic results such that $\mathrm{d} E / \mathrm{d} z$ is continuous (determined numerically). It is assumed that: for incident hard gluons the energy loss scales simply by a factor of $C_A / C_F = 9 / 4$; and that there are enough elastic collisions such that the central limit theorem is applicable, following the WHDG model \cite{Wicks:2005gt}. The latter assumption implies that the distribution of elastic energy loss is Gaussian with mean provided by the BT energy loss formula, and width by the fluctuation dissipation theorem \cite{Moore:2004tg}
\begin{equation}
    \sigma = \frac{2}{E} \int \mathrm{d}z \; \frac{\mathrm{d} E}{\mathrm{d}z} T(z),
    \label{eqn:sigma_elastic}
\end{equation}
where $z$ integrates along the path of the parton, and $T(z)$ is the temperature along the path.

Thus
\begin{multline}
  P_{\text{el}}(p_f | p_i,\,L,\,T) \\ \equiv \frac{1}{\sqrt{2\pi}\sigma}\exp\left[{-}\left( \frac{p_f-(p_i+\Delta p)}{\sqrt{2} \sigma} \right)^2 \right],
\end{multline}
where $\Delta p$ is found by integrating \cref{eqn:elastic_energy_loss_low,eqn:elastic_energy_loss_high} over $z$, and thus
\begin{multline}
  P_{\text{el}}(\epsilon | p_i,\,L,\,T) \\ \equiv \frac{p_i}{\sqrt{2\pi}\sigma}\exp\left[{-}\left( \frac{(1-\epsilon)p_i-(p_i+\Delta p)}{\sqrt{2} \sigma} \right)^2 \right],
\end{multline}
where the additional $p_i$ is the Jacobian resulting from changing variables from $p_f$ to $\epsilon$.

\subsection{Total Energy Loss}%
\label{sec:total_energy_loss}

As done in \cite{Wicks:2005gt}, we convolve the radiative and elastic energy loss probabilities to yield a total probability of energy loss,
\begin{align}
  P_{\text{tot}}(\epsilon) \equiv \int \mathrm{d} x \, P_{\text{el}}(x)P_{\text{rad}}(\epsilon-x).
\end{align}
Note that $P_{\text{rad}}(\epsilon)$ contains Dirac delta functions at both $\epsilon=0$ and $\epsilon=1$ while $P_{\text{el}}(\epsilon)$ is only a Gaussian.  We will see below that the lack of a probability of nothing happening in the elastic energy loss probability is a major shortcoming in modelling the suppression in small collision systems.

\subsection{Geometry}%
\label{sec:geometry}

For large systems it is standard to use the Glauber model for the collision geometry, with Wood-Saxon distributions for the nucleon density inside the heavy ions \cite{Miller:2007ri}. For $\mathrm{p} + \mathrm{A}$ collisions the Glauber model cannot be applied in its most simple form, since subnucleonic features of the proton are expected to be important \cite{Schenke:2020mbo}.  Additionally, the subsequent evolution of the medium can be treated in a more sophisticated way than simply assuming a Bjorken expansion of the initial Glauber model geometry as was done, e.g., in \cite{Wicks:2008zz}.
 In this work we will use collision profiles generated with \cite{Schenke:2020mbo}, and sourced from \cite{shen_private_communication}.
In these calculations, initial conditions are given by the IP-Glasma model \cite{Schenke:2012hg, Schenke:2012wb}, which are then evolved with the \texttt{MUSIC} \cite{Schenke:2010rr,Schenke:2011bn, Schenke:2010nt} viscous relativistic (2+1)D hydrodynamics code, followed by UrQMD microscopic hadronic transport \cite{Bass:1998ca, Bleicher:1999xi}.  In this first comparison, we wish to make as few changes as possible from the original WHDG model and so we will use the initial temperature profile $T(\tau = \tau_0 = 0.4~\mathrm{fm})$ for our collision geometry, where $\tau_0$ is the turn-on time for hydrodynamics.
 This means that we are effectively using the IP-Glasma model \cite{Schenke:2012hg, Schenke:2012wb} as the initial condition coupled with Bjorken expansion time dependence for all presented phenomenological results (unless otherwise stated.)
This method for generating the initial conditions also allows for fluctuating initial conditions, which will be useful for future calculations of azimuthal momentum anisotropy of hard partons $v_n$.

  In addition, future work investigating the use of the complete and realistic time-dependent collision profiles from \cite{Schenke:2020mbo, shen_private_communication} will be of interest.

The QGP is treated as an ultrarelativistic mixture of a Fermi and Bose gas, following \cite{Wicks:2005gt, Horowitz:2010dm}. This results in  standard expressions, for various thermodynamics quantities \cite{Horowitz:2010dm}
\begin{subequations}
    \label{eqn:thermodynamic_quantities}
    \begin{align}
    \mu & = T \sqrt{4 \pi \alpha \left(1 + \frac{n_f}{6}\right)} \\
    \sigma_{g g}  & = \frac{9 \pi \alpha_s^2}{2 \mu^2} \quad \text{and} \quad \sigma_{q g}=\frac{4}{9} \sigma_{g g}\\
    \rho_g & = 16 \frac{\zeta(3)}{\pi^2} T^3 \quad \text{and} \quad  \rho_q = 9 n_f \frac{\zeta(3)}{\pi^2} T^3 \label{eqn:rho_thermal}\\
    \rho & = \rho_g + \frac{\sigma_{q g}}{\sigma_{gg}} \rho_{q} = 4 \frac{\zeta(3)}{\pi^2} T^3(4+ n_f)\label{eqn:color_weighted_density}\\
    \lambda_g^{-1}  & = \rho_g \sigma_{g g}+\rho_q \sigma_{q g} = \sigma_{gg} \rho \label{eqn:mean_free_path}
    \end{align}
\end{subequations}
where $\zeta$ is the Riemann zeta function, $T$ is the temperature, $\rho_q$ ($\rho_g$) is the density of quarks (gluons), $n_f$ is the number of active quark flavors (taken to be $n_f = 2$ throughout), 

$\sigma_{qg}$ ($\sigma_{gg}$) is the gluon-gluon (quark-gluon) cross section, and we have used $N_c = 3$. The cross section weighted density is denoted as $\rho$, which we will subsequently refer to as the density for simplicity.

The radiative (\cref{eqn:full_dndx}) and elastic (\cref{eqn:elastic_energy_loss_low,eqn:elastic_energy_loss_high}) energy loss results were derived using a ``brick" model, which represents a medium with a fixed length $L$ and constant temperature $T$. In order to capture fluctuations in temperature and density, we need a mapping from the path that a parton takes through the plasma, to a brick with an effective length $L_{\text{eff}}$ and effective temperature $T_{\text{eff}}$.

We follow WHDG \cite{Wicks:2005gt} and define the effective pathlength as
\begin{equation}
  L_{\text{eff}} (\mathbf{x}_i, \boldsymbol{\hat{\phi}}) = \frac{1}{\rho_{\text{eff}}} \int \mathrm{d}z \; \rho(\mathbf{x}_i + z \boldsymbol{\hat{\phi}}, \tau_0),
    \label{eqn:effective_length}
\end{equation}
and the effective density as
\begin{equation}
  \rho_{\text{eff}} \equiv \frac{\int \mathrm{d}^2 \mathbf{x} \; \rho^2(\mathbf{x}, \tau_0)}{\int \mathrm{d}^2 \mathbf{x} \; \rho(\mathbf{x}, \tau_0)} \iff T_{\text{eff}}^{3} \equiv \frac{\int \mathrm{d}^2 \mathbf{x} \; T^6(\mathbf{x}, \tau_0)}{\int \mathrm{d}^2 \mathbf{x} \; T^3(\mathbf{x}, \tau_0)}.
  \label{eqn:effective_density}
\end{equation}
Here, the effective pathlength $L_{\text{eff}}$ includes all $(\mathbf{x}_i, \phi)$ dependence, and $\rho_{\text{eff}}$ is a constant for all paths that a parton takes through the plasma for a fixed centrality class. In principle, one can allow both the effective density and effective pathlength to depend on the specific path taken through the plasma.  However, such a numerically intensive model is beyond the scope and objective of this work. Nonetheless, we will discuss some implications of the details of the geometry modelling in \cref{sec:discussion}.

In WHDG \cite{Wicks:2005gt} the prescription $\rho \equiv \rho_{\text{part}}$ was made\footnote{Note that the difference between the density and cross section weighted density falls out at the level of $L_{\text{eff}}$ and $\rho_{\text{eff}}$}, where $\rho_{\text{part}}$ is the participant density---the density of nucleons which participate in at least one binary collision. This prescription is not necessary in our case, since we have access to the temperature profile \cite{Schenke:2020mbo, shen_private_communication}. The temperature is extracted from the hydrodynamics output \cite{Schenke:2020mbo,shen_private_communication}, and for $L_{\mathrm{eff}}$ one evaluates the temperature at the initial time set by the hydrodynamics simulation, $\tau_0=0.4~\mathrm{fm}$. 
There is no unique mapping from realistic collision geometries to simple brick geometries, and more options are explored in \cite{Wicks:2008zz}.

Bjorken expansion \cite{Bjorken:1982qr} is then taken into account by approximating
\begin{equation}
T_{\text{eff}}(\tau) \approx T_{\text{eff}}(\tau_0) \left( \frac{\tau_0}{\tau} \right)^{1/3} \approx T_{\text{eff}}(\tau_0) \left( \frac{2 \tau_0}{L_{\text{eff}}} \right)^{1/3}
  \label{eqn:bjorken_expansion}
\end{equation}
where in the last step we have evaluated $T(\tau, \mathbf{x})$ at the average time $\tau=L / 2$, following what was done in \cite{Wicks:2005gt, Djordjevic:2005db, Djordjevic:2004nq}. In \cite{Wicks:2005gt} this was found to be a good approximation to the full integration through the Bjorken expanding medium.

\begin{figure}[H]
    \centering
    \includegraphics[width=\linewidth]{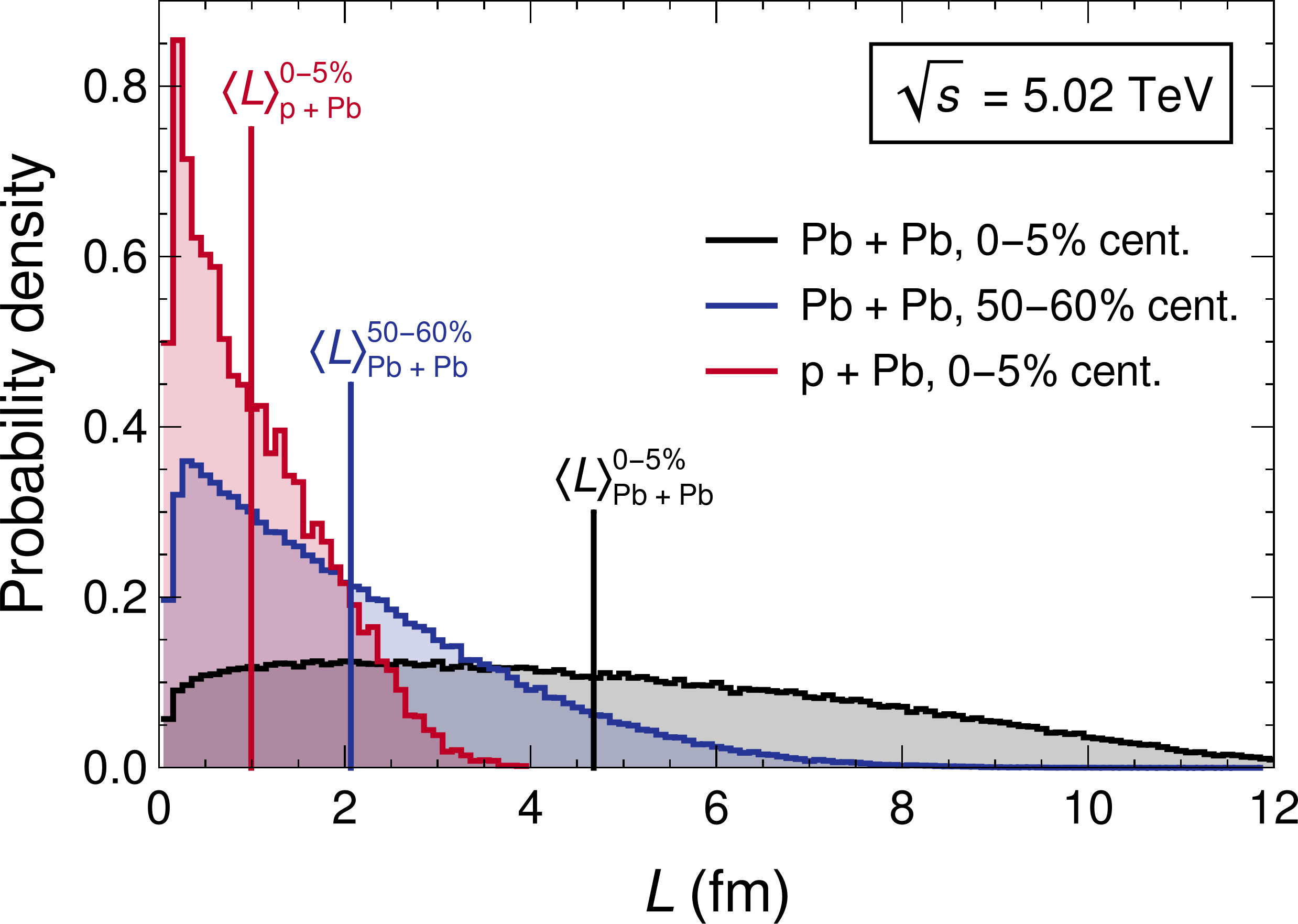}
    \caption{Distribution of the effective pathlengths $P_L(L)$ in $\mathrm{p}+\mathrm{A}$ and $\mathrm{A}+\mathrm{A}$ collision systems, weighted by the binary collision density. The collision systems $\mathrm{Pb}+\mathrm{Pb}$ at 0--5\% and 50--60\% centrality; as well as $\mathrm{p} + \mathrm{Pb}$ at 0--5\% centrality are shown, all at $\sqrt{s} = 5.02~\mathrm{TeV}$. Vertical lines indicates the average lengths for the respective collision systems, which numerically are $0.99 ~\mathrm{fm}$, $2.06~\mathrm{fm}$, and $4.68 ~\mathrm{fm}$ left to right.}
    \label{fig:path_length_distribution}
\end{figure}

For a given collision system we can then calculate the distribution of effective pathlengths that a hard parton will travel in the plasma. We assume, as is standard and consistent with WHDG \cite{Wicks:2005gt}, that the hard partons have starting positions weighted by the density of binary nucleon-nucleon collisions, provided by IP-Glasma \cite{shen_private_communication}. \Cref{fig:path_length_distribution} shows the distribution of effective pathlengths $P_L$ for central $\mathrm{p}+\mathrm{Pb}$ (red), semi-central $\mathrm{Pb}+\mathrm{Pb}$ (blue), and central $\mathrm{Pb}+\mathrm{Pb}$ collisions (black) for $\sqrt{s} = 5.02$ TeV.  We also indicate the average effective pathlength in these three systems by single vertical lines.  One can see that all three distributions are broad, the $\mathrm{p}+\mathrm{Pb}$ system perhaps surprisingly so.  One may further see that the average effective pathlength for central $\mathrm{p}+\mathrm{Pb}$ collisions is not particularly small, $\sim1$ fm, which is not too different from the average effective pathlength for semi-central $\mathrm{Pb}+\mathrm{Pb}$ collisions, $\sim 2$ fm.

\begin{figure}[H]
    \centering
    \includegraphics[width=\linewidth]{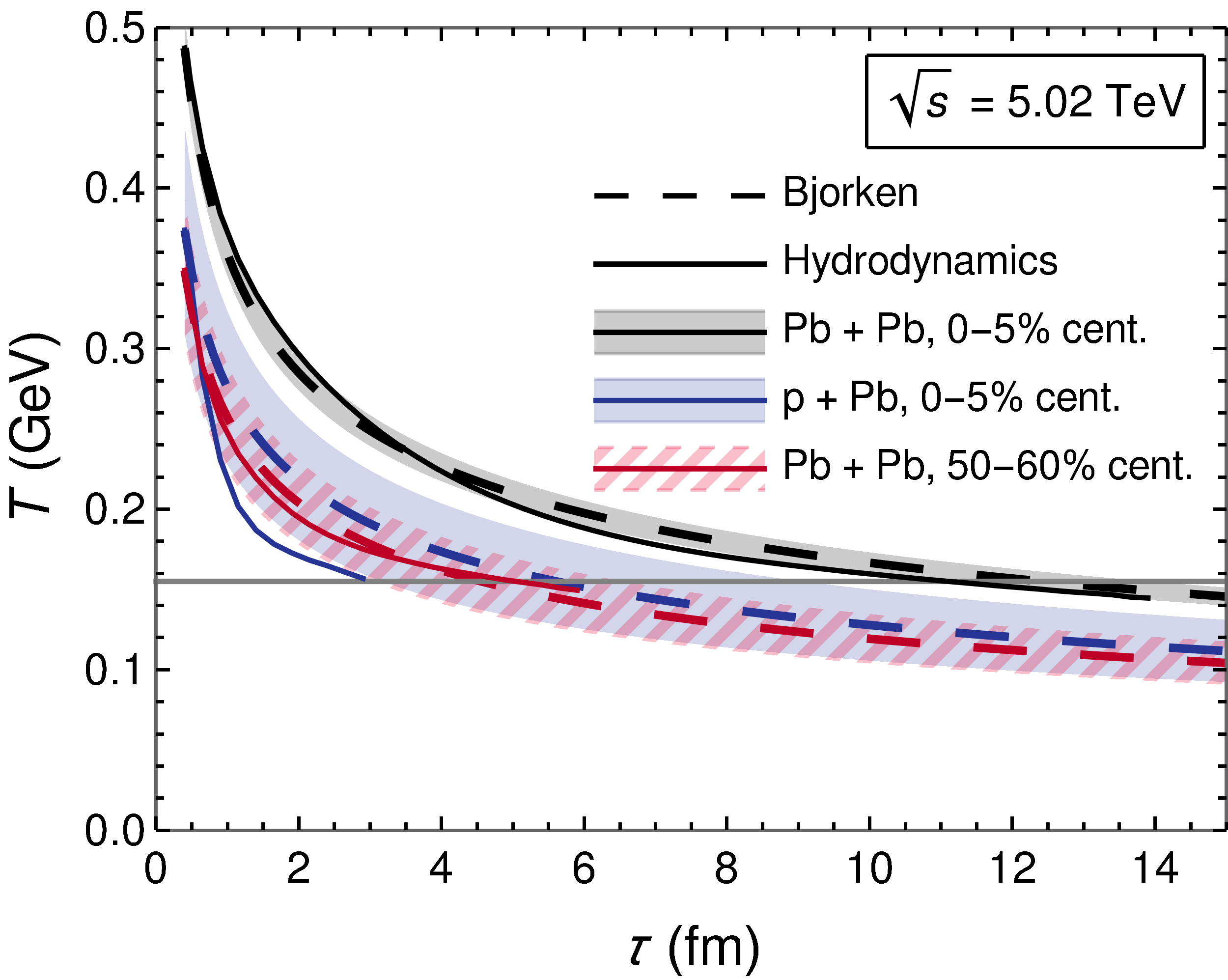}
    \caption{Plot of the temperature $T$ as a function of the proper time $\tau$ (in the plasma rest frame). The Bjorken expansion approximation (Bjorken) \cite{Bjorken:1982qr} to the $\tau$ dependence of temperature, $T(\tau) \approx T_{\text{eff}}(\tau_0) [\tau_0 / \tau]^{1 / 3}$, is plotted along with the effective temperature $T_{\text{eff}}(\tau)$ calculated as a function of time via hydrodynamics (Hydrodynamics) according to \cref{eqn:effective_density}. Uncertainty bands represent a $2 \sigma$ (95\% CI), and are shown only for the Bjorken estimate only as the uncertainty on the Hydrodynamics result is negligible.
    Both curves are plotted for the collision systems $\mathrm{Pb}+\mathrm{Pb}$ at 0--5\%, and 50--60\% centrality; as well as $\mathrm{p} + \mathrm{Pb}$ at 0--5\% centrality. The freeze-out temperature $T_{\text{f.o.}} = 0.155~\mathrm{GeV}$ is shown as a horizontal gray line, which is the temperature at which it is assumed that hadronic degrees of freedom take over in the plasma. Hydrodynamic temperature profiles are taken from \cite{Schenke:2020mbo}. }
    \label{fig:temperature_vs_t}
\end{figure}

\Cref{fig:temperature_vs_t} compares the temperature of the plasma as a function of proper time in the rest frame of the plasma calculated via hydrodynamics (solid lines) versus the temperature from the Bjorken expansion formula (dashed lines). The effective temperature using hydrodynamics is calculated using \cref{eqn:effective_density} and the Bjorken expansion approximation to the time dependence of the effective temperature is given by \cref{eqn:bjorken_expansion}.

  Calculations are performed for the same three collision systems as in \cref{fig:path_length_distribution}. Due to the fluctuations of the initial conditions of the plasma in our model---because of both nucleonic and subnucleonic fluctuations \cite{Schenke:2020mbo}---we obtain a distribution of effective temperatures at each point in proper time. We show the mean and the $2\sigma$ width of the Bjorken temperature estimates in \cref{fig:temperature_vs_t} .  The width in the Bjorken result arises solely from the variation of the initial hydrodynamics temperature profile at $\tau = \tau_0$.  We computed the mean and standard deviation of the temperature distribution as a function of $\tau$ for the full hydrodynamics simulation; however the widths of these distributions are not plotted in \cref{fig:temperature_vs_t} as for $\tau>\tau_0$ they are negligible compared to the width in the Bjorken expansion results.

One can see that for the $\mathrm{A}+\mathrm{A}$ collisions, the Bjorken formula does a very good job of approximating the temporal evolution of the temperature from the full hydrodynamics simulations.  For the central $\mathrm{p}+\mathrm{A}$ system, however, the Bjorken formula significantly overestimates the temperature as a function of time.  We comment below on the effect of this overestimation on the energy loss model. 

The overestimation of the temperature in $\mathrm{p}+\mathrm{A}$ collisions from the Bjorken expansion formula implies that the effective temperature used in the energy loss model is overestimated; as a result, the effects of energy loss are all overestimated in the small collision system.  However, we expect this overestimation to be a small effect; using $\langle L_{\mathrm{eff}}\rangle\sim1$ fm, one can see in \cref{fig:temperature_vs_t} that the difference between the full hydrodynamics temperature and the Bjorken estimate at $\langle \tau \rangle\sim0.5$ fm is very small. Exploring the effect of dynamical effective pathlengths and time-dependent temperatures, which take medium expansion into account using the full hydrodynamic temperature profiles as a function of time, is left for future work.

The average pathlength in 0--5\% most central $\mathrm{p} + \mathrm{Pb}$ collisions is $L\sim 1~\mathrm{fm}$, which has an average temperature of $T \approx 0.35~\mathrm{GeV}$, and correspondingly a mean free path of $\lambda_g \approx 0.75~\mathrm{fm}$.  As one can see from \cref{fig:path_length_distribution}, the large pathlength assumption thus breaks down for the vast majority of hard scatterings in central $\mathrm{p}+\mathrm{A}$ collisions.  Further, in these small collision systems $L/\lambda_g\sim1$ for most of the distribution of effective pathlengths, which implies that approaches that assume many soft scatterings are likely inapplicable.

We note that although the hydrodynamics simulation turns off the QGP phase at the freeze-out temperature $T_{\text{f.o.}} = 0.155~\mathrm{GeV}$, the Bjorken expansion formula \cref{eqn:bjorken_expansion} has no such turn off.  High-$p_T$ particles can of course interact with matter in the hadronic phase, which is, in part, captured by using the Bjorken expansion formula for determining the effective temperature as an input into our energy loss model.  It is worth emphasizing that the effective pathlengths are determined at the initial time $\tau_0=0.4$ fm, well before much of the plasma has had a chance to cool down; thus we do not lose any contribution to the effective pathlengths from hadronization. 

\subsection{Nuclear modification factor}
\label{sub:nuclear_modification_factor}

The observable which we will be computing is the nuclear modification factor $R_{AB}(p_T)$ for a collision system $A+B$, defined \emph{experimentally} by 
\begin{equation}
    R^h_{AB}(p_T) \equiv \frac{\mathrm{d} N^{AB \to h} / \mathrm{d} p_T}{\langle N_{\text{coll}} \rangle \mathrm{d} N^{pp \to h} / \mathrm{d} p_T},
    \label{eqn:nuclear_modification_factor}
\end{equation}

where $\mathrm{d} N^{AB / pp \to h} / \mathrm{d} p_T$ is the differential number of measured $h$ hadrons in $\mathrm{A}+\mathrm{B} / \mathrm{p}+\mathrm{p}$ collisions, and $\langle N_{\text{coll}} \rangle$ is the expected number of binary collisions (usually calculated according to the Glauber model \cite{Miller:2007ri}.)
To access this observable \emph{theoretically} we make several assumptions about the underlying quark and gluon partons.  In the following we will only refer to quarks, but all assumptions and formulae apply equally well for gluons.  We first assume, following \cite{Wicks:2005gt,Horowitz:2010dm}, that the spectrum of produced quarks $q$ in the initial state of the plasma (before energy loss) is $\mathrm{d} N^{q}_{\text{prod}} / \mathrm{d} p_i = N_{\text{coll}} \times \mathrm{d}N^q_{pp} / \mathrm{d}p_i$ where $\mathrm{d} N^q_{pp} / \mathrm{d} p_i$ is the quark production spectrum in $\mathrm{p}+\mathrm{p}$ collisions.
We further assume that the parton production spectra can be approximated by a power law,
\begin{equation}
    \frac{\mathrm{d} N_{pp}^q}{\mathrm{d} p_i}\left(p_i\right) =  \frac{A}{p_i^{n\left(p_i\right)}},
    \label{eqn:slowly_varying_power_law}
\end{equation}
where $n(p_i)$ is slowly varying, and $A$ is a proportionality constant. The $n(p_i)$ function is fitted using the initial parton spectra according to \cref{eqn:slowly_varying_power_law}. For charm and bottom quarks, the initial parton spectra are computed using FONLL\footnote{Practically, the spectra are generated with the online tool at \href{http://www.lpthe.jussieu.fr/~cacciari/fonll/fonllform.html}{http://www.lpthe.jussieu.fr/$\sim$cacciari/fonll/fonllform.html} and then fit to: $\propto p_T^{-n(p_t)}$.} at next-to-leading order \cite{Cacciari:2001td}; and for gluons and light quarks, production spectra are computed\footnote{Note that the fraction of light quarks to gluons is obtained from Fig.\ 2 in \cite{Horowitz:2011gd}.} to leading order \cite{wang_private_communication} as in \cite{Vitev:2002pf, Horowitz:2011gd}. 

From \cref{sec:total_energy_loss}, we have the total probability density function $P_{\text{tot}}(\epsilon | p_i)$, which is the probability for a parton with initial transverse momentum $p_i$ to lose a fraction $\epsilon$ of it's energy such that the final momentum is $p_T = (1 - \epsilon) p_i$. Assuming that particle spectra are modified primarily due to energy loss implies,
\begin{equation}
  \mathrm{d} N^q\left(p_T\right)=\mathrm{d} N^q\left(p_i\right) P_{\text{tot}}\left(\epsilon \mid p_i\right) \mathrm{d} \epsilon,
  \label{eqn:probability_energy_loss_and_differential_number}
\end{equation}
where $\mathrm{d} N^{q}(p_{i})$ ($\mathrm{d} N^{q}(p_{T})$) is the differential number of quarks in the initial (final) state with momentum $p_i$ ($p_T$). Finally we assume that $P_{\text{tot}}(\epsilon | p_i)$ varies slowly with $p_i$, leading to the following expression for the partonic $R^{q}_{AB}$ (neglecting hadronization for now) \cite{Horowitz:2010dm}
\begin{align}
  R_{A B}^q\left(p_T\right) \equiv& \frac{\mathrm{d} N^{AB \to q} / \mathrm{d} p_T}{N_{\text{coll}} \times \mathrm{d} N^{pp \to q} / \mathrm{d} p_T}\\
  =& \frac{N_{\mathrm{coll}} \int \frac{\mathrm{d} \epsilon}{1-\epsilon} \frac{A}{\left(p_T / 1-\epsilon\right)^{n\left(p_T / [1-\epsilon]\right)}} P\left(\epsilon \mid \frac{p_T}{1-\epsilon}\right)}{N_{\mathrm{coll}} \; A \; p_T^{- n\left(p_T\right)}}\label{eqn:simple_raa_before_slowly_varying}\\
  \simeq& \int \mathrm{d} \epsilon \; P_{\text{tot}}\left(\epsilon \mid p_T\right)(1-\epsilon)^{n\left(p_T\right)-1}.
    \label{eqn:simple_raa}
\end{align}
In the above we have used $P\left(\epsilon | p_T / [1-\epsilon]\right) \approx P(\epsilon | p_T)$ and $n[p_T / (1-\epsilon)] \approx n(p_T)$ which follows from the slowly varying assumptions about $n(p_T)$ and $P(\epsilon | p_T)$ and the soft assumption $\epsilon \ll 1$. The assumption that $n(p_T)$ varies slowly, can easily be verified when fitting \cref{eqn:slowly_varying_power_law} to spectra. 
Since radiative energy loss is dominant, $P(\epsilon | p_T)$ will be peaked at $\Delta E / E$ determined with \cref{eqn:fractional_energy_loss}. Asymptotically this grows as $\log (E) / E$  for the (D)GLV result \cite{Gyulassy:2000er}, and as $\log (E)$ for the correction \cite{Kolbe:2015rvk}. It is safe to assume this logarithmic growth is slow, however at high energies where the short pathlength correction dominates this assumption may become worse.

To incorporate our model for the collision geometry, we expand \cref{eqn:simple_raa} to average over the effective pathlengths according to the length distribution described in \cref{sec:geometry},
\begin{align}
  R&_{A B}^q\left(p_T\right) \nonumber\\
  & =  \langle R_{A B}^{q}(p_T, L, T) \rangle_{\text{geometry}}\nonumber\\
  & = \int \mathrm{d} L_{\mathrm{eff}}\; P_L(L_{\mathrm{eff}}) \int \mathrm{d} \epsilon \label{eqn:geometry_averaged_raa}\\
  & \quad \times P_{\text{tot}}\left(\epsilon \mid \{p_T, L_{\mathrm{eff}}, T (L_{\text{eff}} / 2)\}\right)(1-\epsilon)^{n\left(p_T\right)-1},\nonumber
\end{align}
where $P_L(L_{\mathrm{eff}})$ is the normalized distribution of effective pathlengths weighted weighted by the binary collision density (\cref{fig:path_length_distribution}) and $T(L_{\text{eff}} / 2)$ is the effective temperature determined according to \cref{eqn:bjorken_expansion}.

The spectrum $\mathrm{d} N^h / \mathrm{d} p_h$ for a hadron $h$ is related to the spectrum $\mathrm{d} N^q / \mathrm{d} p_q$ for a parton $q$ via \cite{Horowitz:2010dm}
\begin{equation}
        \frac{\mathrm{d} N^h}{\mathrm{d} p_h}\left(p_h\right) =\int \frac{\mathrm{d} N^q}{\mathrm{d} p_q}\left(\frac{p_h}{z}\right) \frac{1}{z} D^h_q(z, Q) \mathrm{d} z,
    \label{eqn:parton_to_hadron_spectrum}
\end{equation}
where $z = p_q / p_h \in (0,1]$, $p_h$ is the observed hadron momentum, $D^h_q(z,Q)$ is the fragmentation function for the process $q \mapsto h$, and $Q$ is the hard scale of  the problem taken to be $Q = p_h / z$. The hadronic $R^h_{AB}$ is then found in terms of the partonic $R^q_{AB}$ (\cref{eqn:geometry_averaged_raa}) as \cite{Horowitz:2010dm}
\begin{align}
    R_{AB}^h\left(p_T\right)&=\frac{\sum_q \int d z \frac{1}{z} D_q^h(z) \frac{d N_{pp}^q}{d p_q}\left(\frac{p_T}{z}\right) R_{AB}^q\left(\frac{p_T}{z}\right)}{\sum_q \int d z \frac{1}{z} D_q^h(z) \frac{d N_{p p}^q}{d p_q}\left(\frac{p_T}{z}\right)}.
    \label{eqn:parton_to_hadron_raa}
\end{align}
For details of the derivation needed for both \cref{eqn:simple_raa,eqn:parton_to_hadron_raa}, refer to Appendix B of \cite{Horowitz:2010dm}.
The fragmentation functions for $D$ and $B$ mesons are taken from \cite{Cacciari:2005uk}, and the fragmentation functions for $\pi$ mesons are taken from \cite{deFlorian:2007aj}. Note that the fragmentation functions for $\pi$ mesons were extrapolated outside their domain in $Q^2$, as we found that the extrapolation was smooth.
There are uncertainties in both the fragmentation function fits and the generation of initial parton spectra's; however these uncertainties are very small compared to others in this problem (running coupling, first order in opacity, etc.), and so we will not take the fragmentation function uncertainties into account.

\section{Initial results}
\label{sec:results}

As a first exploration of the effect of the short pathlength correction to the DGLV radiative energy loss model, we consider $\mathrm{Pb}+\mathrm{Pb}$ and $\mathrm{p} + \mathrm{Pb}$ collision systems. We emphasize that in this work we are not trying to create the best possible fit to data (which could be done by tuning various parameters), but are instead focused on the impact of the short pathlength correction. For this reason all numerical values are used in consistency with the original WHDG predictions \cite{Wicks:2005gt,Horowitz:2011gd,Andronic:2015wma}, and agreement with data is not the primary focus of this work.

\subsection{Suppression of heavy flavor mesons}
\label{sub:results_heavy}

\begin{figure}[H]
    \centering
    \includegraphics[width=\linewidth]{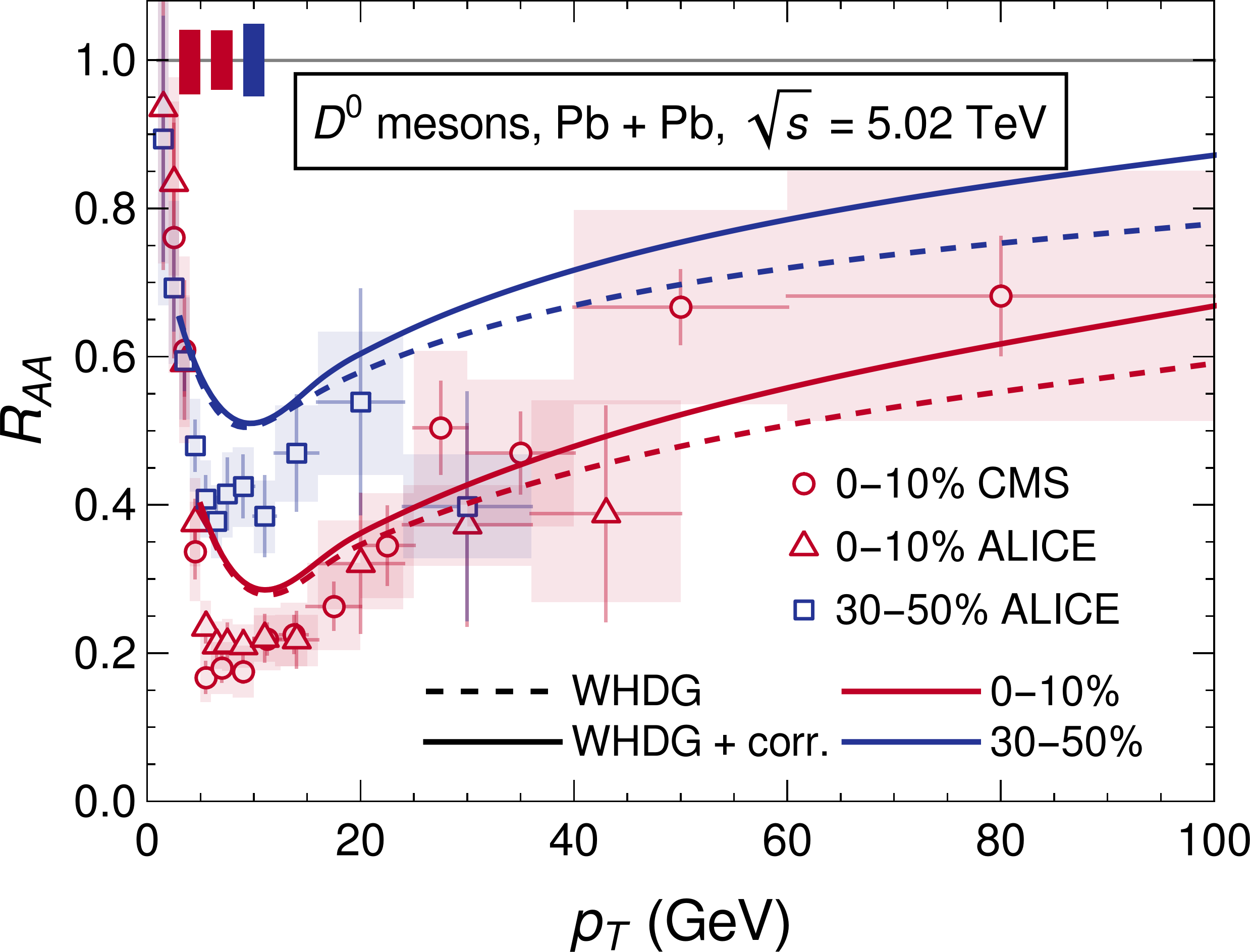}
    \caption{ The nuclear modification factor $R_{AA}$ as a function of final transverse momentum $p_T$ is calculated for $D^0$ mesons, with and without the short pathlength correction to the radiative energy loss. Calculations were done for 0--10\% centrality as well as 30--50\% centrality. Data from CMS \cite{CMS:2017qjw}, and ALICE \cite{ALICE:2018lyv} are shown for comparison; where error bars (boxes) indicate statistical (systematic) uncertainties. The global normalisation uncertainty on the number of binary collisions is indicated by the solid boxes in the top left corner of the plot (left to right: 0--10\% CMS, 0--10\% ALICE, 30--50\% ALICE).}
    \label{fig:D_raa_AA}
\end{figure}

\begin{figure}[H]
    \centering
    \includegraphics[width=\linewidth]{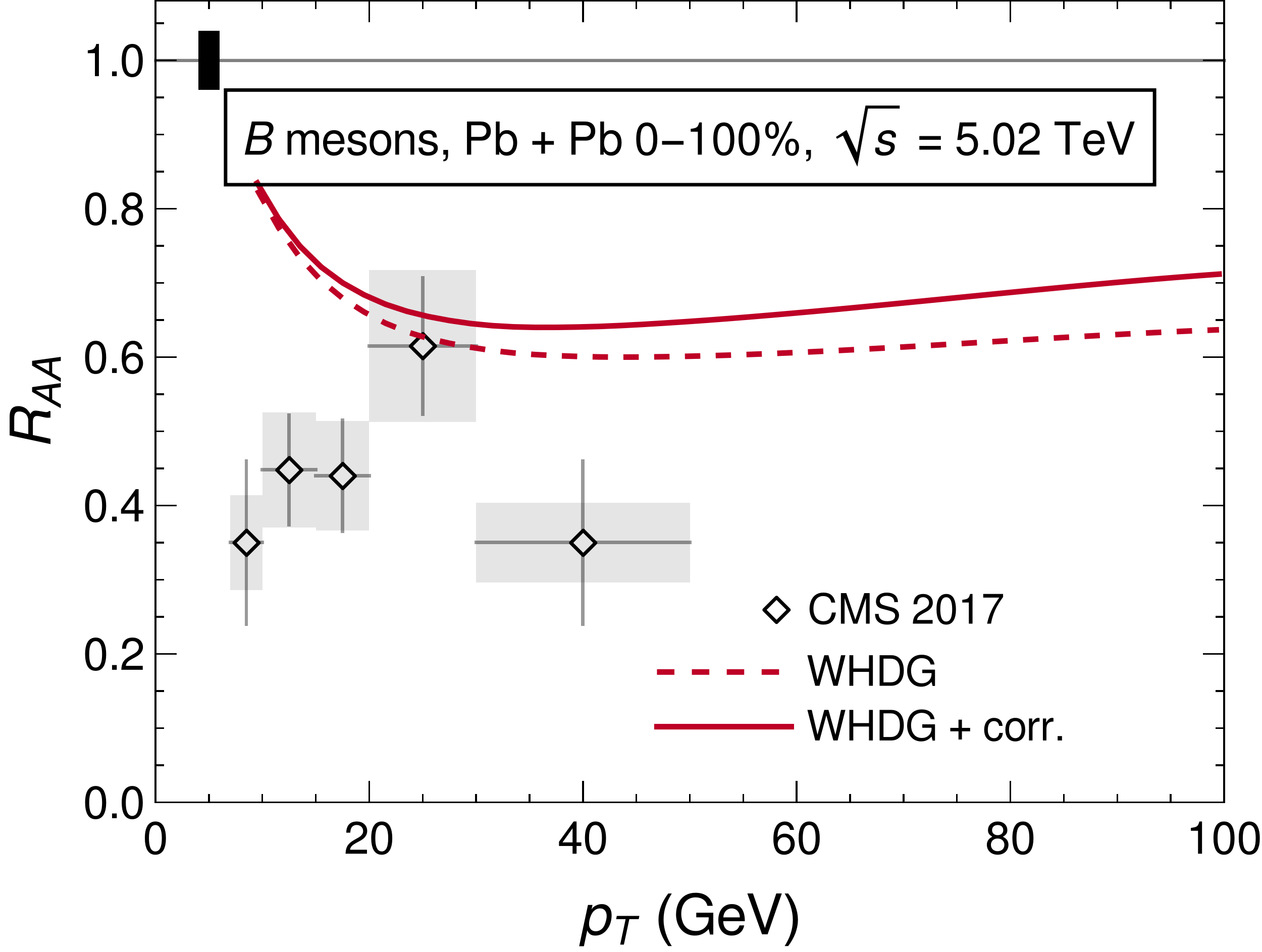}
    \caption{ The nuclear modification factor $R_{AA}$ as a function of final transverse momentum $p_T$ is calculated for $B$ mesons with and without the short pathlength correction. Data for $B^\pm$ mesons from CMS \cite{CMS:2017uoy} is shown for comparison, where error bars (boxes) indicate statistical (systematic) uncertainties. The global normalisation uncertainty on the number of binary collisions is indicated by the solid boxes in the top left corner of the plot.}
    \label{fig:B_raa_AA}
\end{figure}

In \cref{fig:D_raa_AA}, we show $R_{AA}(p_T)$ for $D$ mesons in $\sqrt{s} = 5.02$ TeV $\mathrm{Pb}+\mathrm{Pb}$ collisions at 0--10\% and 30--50\% centrality from our convolved radiative and collisional energy loss model, with and without the short pathlength correction to DGLV radiative energy loss, compared to data from CMS \cite{CMS:2017qjw}, and ALICE \cite{ALICE:2018lyv}.  In \cref{fig:B_raa_AA}, we show $R_{AA}(p_T)$ for $B$ mesons in $\sqrt{s} = 5.02$ TeV $\mathrm{Pb}+\mathrm{Pb}$ collisions at 0--10\% centrality from our convolved radiative and collisional energy loss model, with and without the short pathlength correction to DGLV radiative energy loss, compared to minimum bias data from CMS \cite{CMS:2017uoy}.  

From the figures we conclude that for heavy flavor final states, the short pathlength correction to the $R_{AA}$ is small up to $p_T \sim 100~\mathrm{GeV}$, and that the difference grows with $p_T$. This small difference was expected from the numerical energy loss calculations performed in \cite{Kolbe:2015rvk} and reproduced in \cref{fig:deltaEOverEShortvsDGLV} that showed the small pathlength corrections are a relatively small correction for quarks. Agreement with data for $D^0$ mesons (\cref{fig:D_raa_AA}) is especially good for all $p_T \gtrsim 5~\mathrm{GeV}$, given that the calculation is leading order. The $D^0$ meson suppression prediction is underestimated for moderate $p_T \lesssim 20~\mathrm{GeV}$ in comparison to a previous prediction with the WHDG model in \cite{Andronic:2015wma}. We believe the main cause of this difference with past results is from not using the approximation $q_{\text{max}}\to \infty$ (see \cref{sub:radiative_energy_loss}), which overestimates the energy loss---especially at low $p_T$.

\begin{figure}[H]
    \centering
    \includegraphics[width=\linewidth]{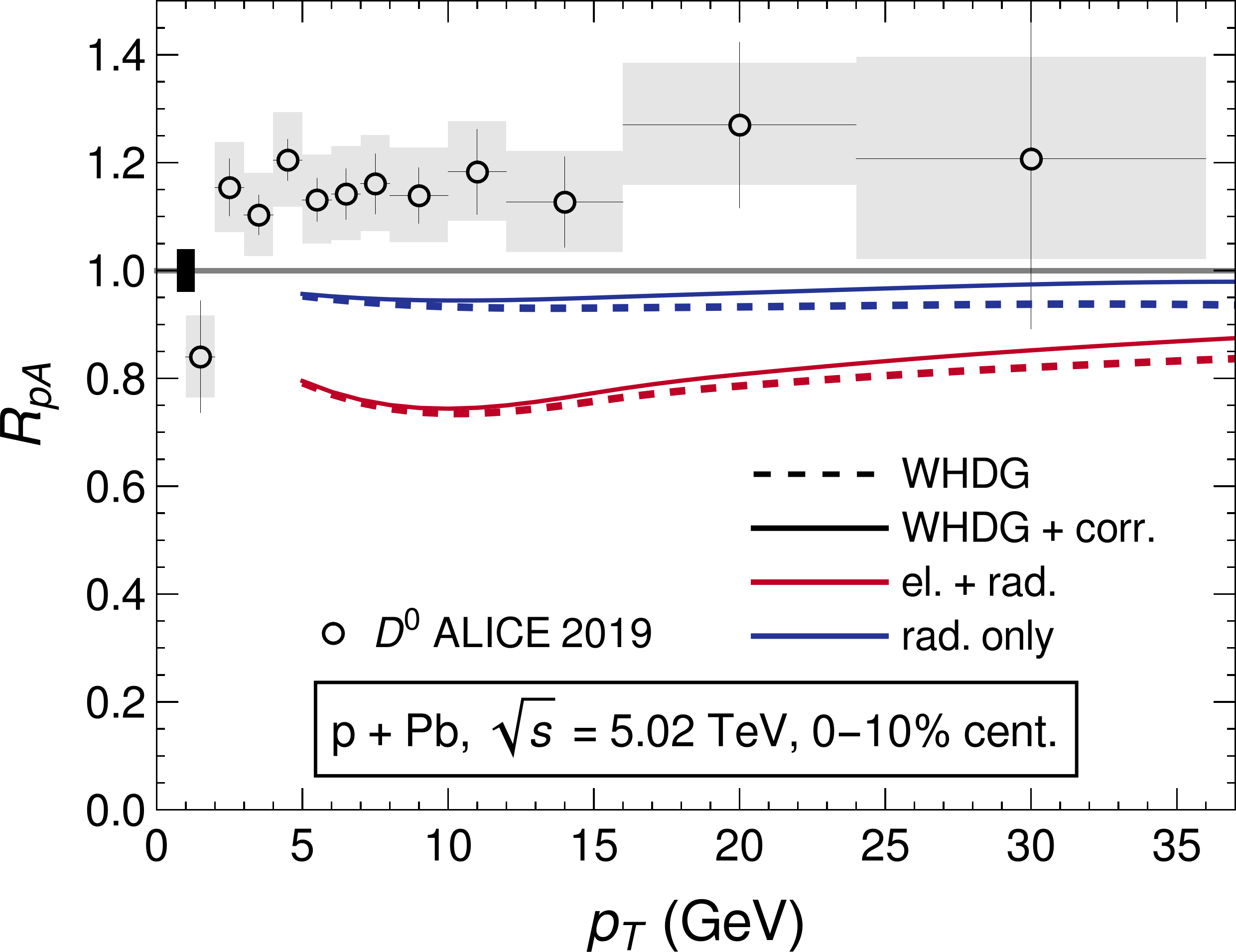}
    \caption{ The nuclear modification factor $R_{pA}$ for $D^0$ mesons as a function of final transverse momentum $p_T$ is calculated with and without the short pathlength correction. The $R_{pA}$ is calculated both with collisional and radiative energy loss (el.\ + rad.), and with radiative energy loss only (rad.\ only). Data from ALICE \cite{ALICE:2019fhe} for $D^0$ mesons is shown for comparison, where statistical (systematic) uncertainties are represented by error bars (boxes). The global normalisation uncertainty on the number of binary collisions is indicated by the solid box in the center left of the plot.}
    \label{fig:D_raa_pA}
\end{figure}

\Cref{fig:D_raa_pA} shows predictions for the $R_{pA}$ of $D$ mesons as a function of the final transverse momentum $p_T$ for central $\mathrm{p} + \mathrm{Pb}$ collisions at $\sqrt{s} = 5.02$ TeV from our convolved radiative and collisional energy loss model compared to ALICE data \cite{ALICE:2019fhe}.  In the same figure, we also show the prediction of $R_{pA}$ of $D$ mesons from our energy loss model with the collisional engery loss turned off.  One can see that the energy loss model that includes both collisional and radiative energy loss, both the small pathlength corrected and the uncorrected version, dramatically overpredicts the suppression in this small system.  At the same time, the predictions from the model with the collisional energy loss turned off are significantly less oversuppressed compared to the data.  The surprising sensitivity to the presence of the collisional energy loss process is due to our naive implementation of the collisional energy loss as discussed in \cref{sec:total_energy_loss}: the WHDG \cite{Wicks:2005gt} collisional energy loss model assumes an average collisional energy loss with Gaussian fluctuations, which is inappropriate for a small system with very few elastic scatterings.

\subsection{Suppression of light flavor mesons}
\label{sub:results_light}

\begin{figure}[H]
    \centering
    \includegraphics[width=\linewidth]{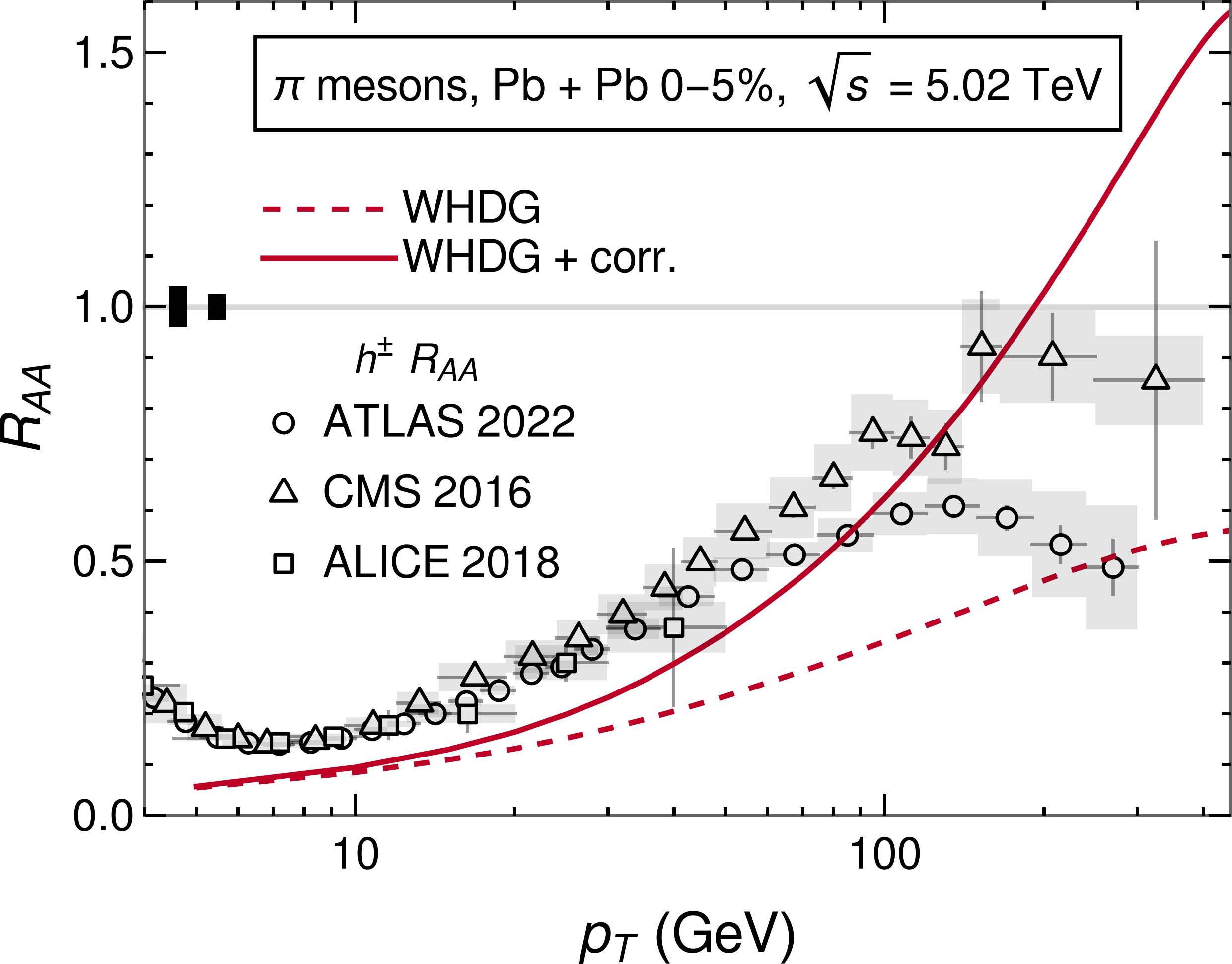}
    \caption{The nuclear modification factor $R_{AA}$ for $\pi$ mesons produced in $\mathrm{Pb}+\mathrm{Pb}$ collisions at 0--5\% centrality is calculated using WHDG, with and without the short pathlength correction to the radiative energy loss. Charged hadron suppression data from ATLAS \cite{ATLAS:2022kqu}, CMS \cite{CMS:2016xef}, and ALICE \cite{Sekihata:2018lwz} is plotted for comparison; with error bars (boxes) corresponding to statistical (systematic) uncertainties. The global normalisation uncertainty on the number of binary collisions is indicated by the solid boxes in the center left of the plot (left to right: CMS, ALICE). The normalisation uncertainty for the ATLAS data is included in the systematic uncertainties.}
    \label{fig:pion_raa_AA}
\end{figure}

\Cref{fig:pion_raa_AA} shows the $R_{AA}(p_T)$ for $\pi$ mesons as a function of $p_T$ in 0--5\% most central $\mathrm{Pb}+\mathrm{Pb}$ collisions at $\sqrt{s} = 5.02$ TeV from both our convolved radiative and collisional energy loss model compared to the $R_{AA}(p_T)$ for charged hadrons measured by ATLAS \cite{ATLAS:2022kqu}, CMS \cite{CMS:2016xef}, and ALICE \cite{Sekihata:2018lwz}. In our energy loss model, we used the fraction of $\pi$ mesons originating from light quarks versus gluons as in \cite{Horowitz:2011gd}. 

\Cref{fig:pion_raa_AA} shows that the short pathlength correction does, in fact, lead to a stronger $p_T$ dependence in $R_{AA}(p_T)$ than the prediction without the correction.  In fact, the correction leads to a large change in predicted $R_{AA}(p_T)$, even at moderate momenta $20~\mathrm{GeV} \lesssim p_T \lesssim 100~\mathrm{GeV}$, when compared to the size of the correction for heavy flavor mesons.  This large change in $R_{AA}(p_T)$ is consistent with the large change in the average energy loss as calculated numerically in \cite{Kolbe:2015rvk} and reproduced in \cref{fig:deltaEOverEShortvsDGLV,fig:deltaE_vs_energy}; the correction is almost a factor of 10 times larger for gluons compared to quarks due to the specific way in which color triviality is broken by the short pathlength correction to the energy loss. For $p_T \lesssim 200~\mathrm{GeV}$ the corrected result is tantalizingly consistent with data, however for $p_T \gtrsim 200~\mathrm{GeV}$ the corrected result predicts anomalously large enhancement up to $R_{AA} = 1.5$ at $p_T \simeq 450~\mathrm{GeV}$, a shocking $\sim \!\! 200\%$ increase over the uncorrected result. 

\begin{figure}[H]
    \centering
    \includegraphics[width=\linewidth]{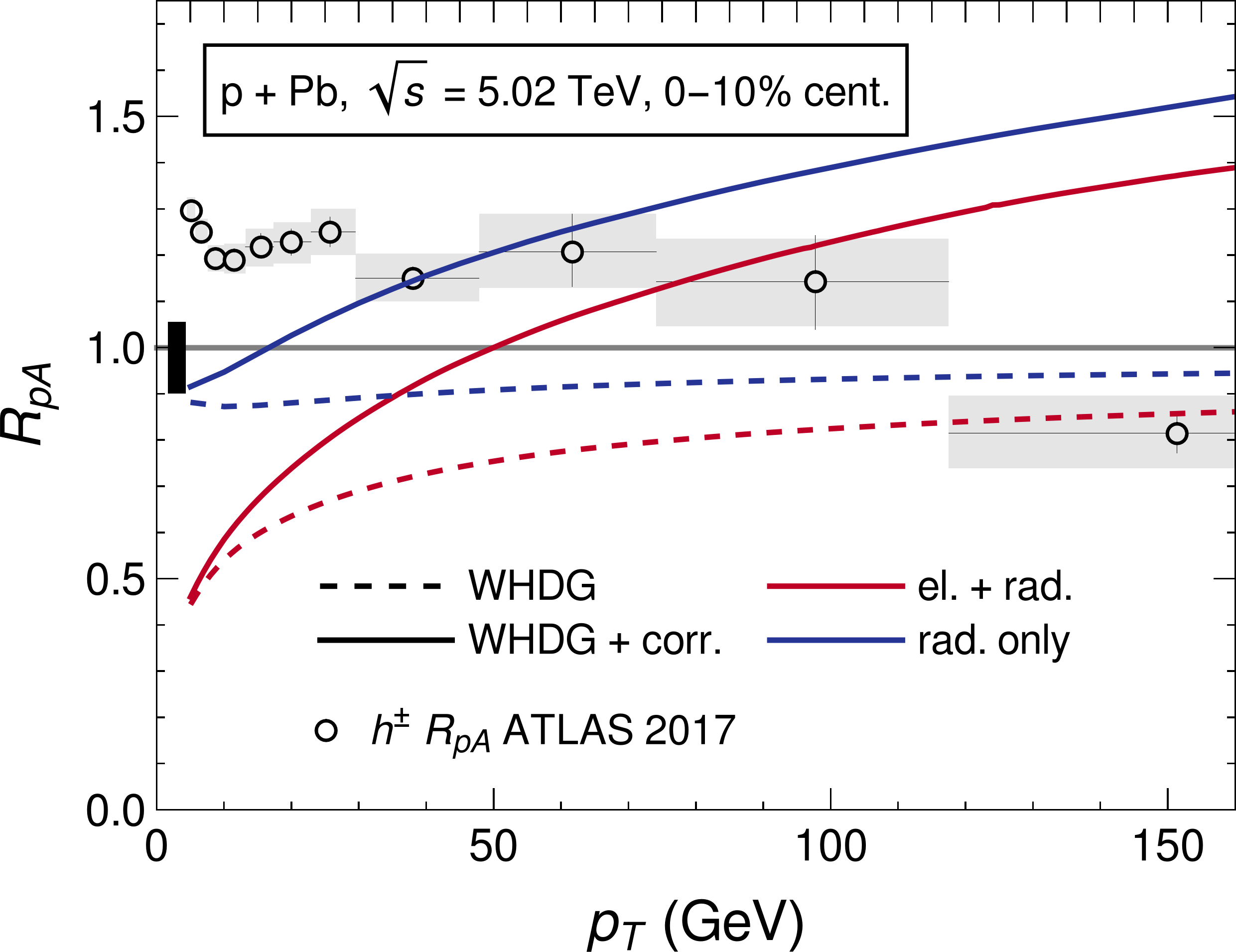}
    \caption{ The nuclear modification factor $R_{pA}$ for $\pi$ mesons as a function of final transverse momentum $p_T$, is calculated with and without the short pathlength correction to the radiative energy loss. The $R_{pA}$ is calculated both with collisional and radiative energy loss (el. + rad.), and with radiative energy loss only (rad. only). Data from ATLAS \cite{Balek:2017man} for charged hadrons is shown for comparison, where statistical (systematic) uncertainties are represented by error bars (boxes). The global normalisation uncertainty on the number of binary collisions is indicated by the solid box in the top left corner of the plot.}
    \label{fig:pion_raa_pA}
\end{figure}

\Cref{fig:pion_raa_pA} shows predictions for the $R_{pA}(p_T)$ of $\pi$ mesons in 0--10\% most central $\mathrm{p} + \mathrm{Pb}$ collisions from our convolved radiative and collisional energy loss model, with and without the short pathlength correction to the radiative energy loss, as well as predictions from our model with the collisional energy loss turned off.  \cref{fig:pion_raa_pA} also shows ATLAS charged hadron $R_{pA}(p_T)$ data \cite{Balek:2017man}. In this figure the breakdown of the elastic energy loss is even more obvious than for the heavy flavor mesons, since the elastic energy loss for gluons is enhanced by a factor $C_A / C_F \approx 2$. The uncorrected radiative-only result (no collisional energy loss) predicts mild suppression of $R_{AA} \approx 0.9$ for all $p_T$; and the corrected radiative-only result predicts enhancement that grows monotonically in $p_T$, reaching $R_{AA} \approx 1.5$ at $p_T = 150~\mathrm{GeV}$. The predicted enhancement for $p_T \gtrsim 100~\mathrm{GeV}$ is in excess of the measured enhancement \cite{Balek:2017man}, however for $p_T \lesssim 100~\mathrm{GeV}$ the presence of enhancement is qualitatively consistent with data.

The short pathlength ``correction'' leads to changes in $R_{AA}(p_T)$ of 100\% or greater in the light hadron sector, with predictions of $R_{AA}(p_T)$ well in excess of 1 for extremely high momenta, in both small and large collision systems leads us to question whether or not the energy loss calculation is breaking down in some fundamental way.  In particular, are we applying the energy loss formulae in our energy loss model in some regimes where the assumptions made in the derivation of those energy loss formulae no longer apply?

\section{Consistency of assumptions in DGLV}
\label{sec:assumptions}

The prediction of significant enhancement of high-$p_T$ light flavor mesons shown in \cref{fig:pion_raa_AA,fig:pion_raa_pA} stems from the asymptotic dependence of the short pathlength correction on energy \cite{Kolbe:2015rvk}.  
We see from \cref{eqn:asymptotics_correction,eqn:asymptotics_DGLV} that for asymptotically large values of energy, $\Delta E_{\text{corr.}} / E \sim E^0$ while $\Delta E_{\text{DGLV}} / E\sim \ln E/E$.  Thus, inevitably, the correction becomes larger than the uncorrected result in the large $E$ limit.  Presumably, then, there's some intermediate value of the energy at which the assumptions that went into either the DGLV derivation, the derivation of the correction, or both are violated. As noted in the Introduction, the derivations of both DGLV energy loss and its small pathlength correction assumed: (1) the \textit{Eikonal approximation} which assumes that the energy of the hard parton is the largest scale in the problem; (2) the \textit{soft radiation approximation} which assumes $x \ll 1$; (3) \textit{collinearity} which assumes $k^+  \gg k^-$; (4) the impact parameter varies over a large transverse area; and (5) the \textit{large formation time assumption} which assumes $\omega_{0} \ll \mu \Leftrightarrow \mathbf{k}^2 / 2 x E \ll \mu$ and $\omega_1 \ll \mu_1 \Leftrightarrow (\mathbf{k} - \mathbf{q_1})^2 / 2 x E \ll \sqrt{\mu^2 + q_1^2}$. For the large formation time assumption we found that in the original calculation (details in \cite{Kolbe:2015suq}), $\omega_0 \ll \mu$ was only used in the weaker form $\omega_0 \ll \mu_1$; and so we will be considering this weaker assumption instead.

In this section we numerically check the consistency of these assumptions with the final radiative energy loss result.  In particular, the analytic properties of the matrix element mean that it may have non-zero support for momenta that are unphysical (even complex).  The relevant question for us is: does the matrix element (modulus squared) give a significant contribution to the energy loss in kinematic regions that are integrated over but for which the derivation of the matrix element is not under control? %

In an attempt to (partially) answer this question, we are motivated to calculate expectation values of ratios assumed small under the various assumptions, weighted by the absolute value of the mean radiative energy loss distribution, \cref{eqn:fractional_energy_loss}\footnote{Just to be clear, we are weighting by the mean radiative energy loss as determined by the single inclusive gluon emission distribution, $\sim x \, \mathrm{d} N^g/\mathrm{d}x$; we are not weighting by the Poisson convolved distribution.}. Explicitly the procedure to calculate the expectation value of a function $R(\{ X_i \})$, depending on the set of variables $\{X _i\}$, is
\begin{equation}
  \langle R \rangle \equiv \frac{\int \mathrm{d} \{X_i\} ~ R(\{X_i\}) \; \left | \frac{\mathrm{d} E}{\mathrm{d} \{ X_i \}} \right |}{\int \mathrm{d} \{X_i\}~ \left | \frac{\mathrm{d} E}{\mathrm{d} \{X_i\}} \right |},
    \label{eqn:expected_value}
\end{equation}
where $\{X_i\}$ can be any of $\{\mathbf{k}, \mathbf{q}, x, \Delta z\}$ and $\mathrm{d} \{ X_i \} \equiv \prod_i \mathrm{d} X_i$. Also note that $R$ can depend on quantities that are not integrated over, such as $\{L, E, \mu\}$. It is important to note that this expectation value is not an expectation value in the usual sense, where the distribution is the distribution of radiated gluons, because we are weighting by the radiative energy loss and not radiated gluon number.  It is also important to note that even if a particular assumption is violated in the sense of this weighted average, that violation does not necessarily mean that the correction computed by relaxing the assumption is large; rather, we only know that the correction is not necessarily small.

\begin{figure}[H]
    \centering
    \includegraphics[width=\linewidth]{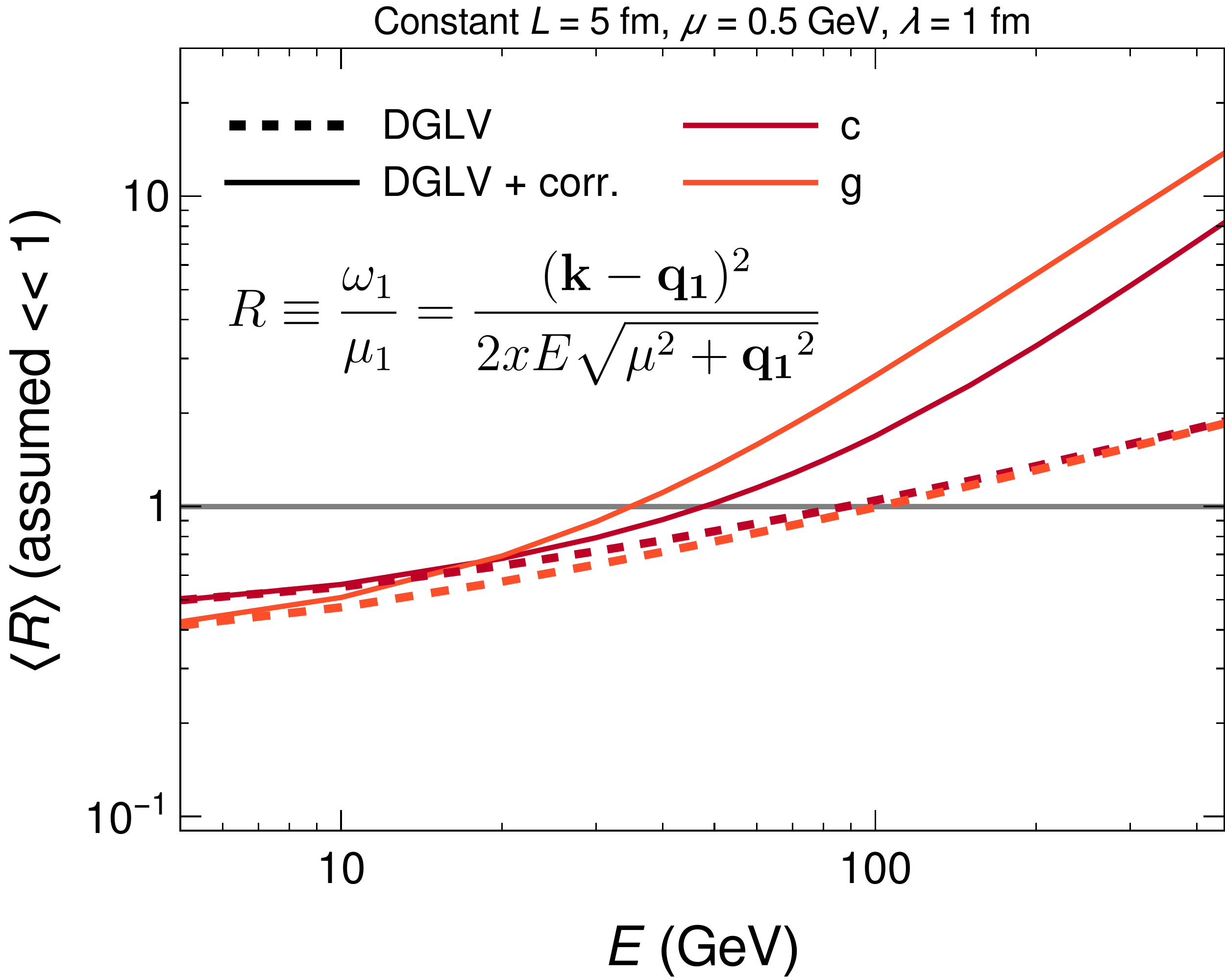}
    \caption{
    Plot of $\langle R\rangle \equiv \langle \omega_1/\mu_1 \rangle$ as a function of parent parton energy $E$.  $\langle R \rangle\ll1$ implies consistency with the large formation time assumption.  $\langle R \rangle$ is computed without (dashed) and with (solid) the short pathlength correction for charm quarks (dark red) and gluons (orange).  
    All curves use constant $L = 5~\mathrm{fm}$, $\lambda_g=1~\mathrm{fm}$, and $\mu=0.5~\mathrm{GeV}$.
  }
    \label{fig:largeformationtime2NoStep}
\end{figure}

\Cref{fig:largeformationtime2NoStep} investigates the large formation time assumption.  The figure shows the expectation value of $\omega_1 / \mu_1 = (\mathbf{k}-\mathbf{q}_1)^2/2xE\sqrt{\mu^2+\mathbf{q}_1^2}$, where the DGLV and correction derivations assume that $\omega_1 / \mu_1 \ll 1$, using $L = 5~\mathrm{fm}$, $\lambda_g=1~\mathrm{fm}$, and $\mu=0.5~\mathrm{GeV}$. 
The large formation time assumption is explicitly violated for the energy loss with and without the short pathlength correction. For the energy loss without the correction, the large formation time assumption is violated for $E \gtrsim 100$ GeV for both charm quarks and gluons, while for the energy loss with the correction the large formation time assumption is violated for $E \gtrsim 35~\mathrm{GeV}$ for gluons and for $E \gtrsim 50~\mathrm{GeV}$ for charm quarks. 
This breakdown calls into question the validity of the large formation time assumption in DGLV radiative energy loss for $p_T \gtrsim 100$ GeV, regardless of whether the energy loss receives a short pathlength correction. The increased rate and magnitude of large formation time assumption violation once the short pathlength correction is included in the energy loss indicates that the short pathlength corrected energy loss model predictions may be breaking down at moderate to high $p_T$ in central $\mathrm{A}+\mathrm{A}$ collisions. One sees that the large formation time assumption, always breaks down before enhancement, $\Delta E / E < 0$, is predicted.

One possibility for the increased rate and degree of large formation time breakdown once the short pathlength is included
---and the erroneously large correction at high $p_T$ for $\pi$ mesons---is the emphasis placed on short pathlengths by the exponential distribution of scattering centers. The exponential distribution was originally chosen to make the analytics simple, with the physical motivation that it captures the Bjorken expansion in the medium \cite{Djordjevic:2003zk}.

Bjorken expansion leads to a power law decay of the plasma density in time \cite{Bjorken:1982qr}, and so an exponential distribution likely overestimates the amount of expansion. This biases scatterings to occur at smaller $\Delta z$ than is physical, and likely overestimates the contribution from the short pathlength correction. Additionally it is not obvious how to model the time dependence of the collision geometry before thermalization $\tau \lesssim \tau_0$, as in principle the medium should be thermalizing during this time. Furthermore the treatment of scatters that occur for times $\tau \lesssim \tau_0$ is not obvious as it is possible that the well-separated scattering centers assumption $\lambda  \gg \mu^{-1}$ breaks down in this phase of the plasma. It was found numerically that DGLV energy loss results are insensitive to the exact distribution of scattering centers used \cite{Armesto:2011ht}.  Perhaps not surprisingly, the small pathlength correction has a large sensitivity to the exact distribution of scattering centers used \cite{Kolbe:2015rvk}.

We are thus motivated to consider an alternative distribution of scattering centers as we consider whether or not the various assumptions made in the energy loss derivations are consistent with our final energy loss numerics.  In this paper we will consider, in addition to the usual exponential distribution, the \textit{truncated step} distribution from \cite{Kolbe:2015rvk}.  The truncated step distribution is given by $\bar{\rho}_{\text{step}}(\Delta z) \equiv (L-a)^{-1} \Theta(\Delta z-a) \Theta(L-\Delta z)$, where $a$ is a small distance cut off.  The truncated step function attempts to capture the effect of a ``turn on'' of the QGP, before which no energy loss takes place, with subsequent equal probability for a scattering to occur until the end of the pathlength.  We think of the exponential distribution and truncated step distributions as limiting cases for what the real distribution of scattering centers may be. The exponential distribution maximally emphasizes the effect of early-time physics, while the truncated step distribution completely neglects early-time physics. A more realistic distribution is likely somewhere in between these two extremes. 

One choice for $a$ is $a = \mu^{-1}$, since for $\Delta z \lesssim \mu^{-1}$, the production point and scattering center are too close to be individually resolved. Another choice is $a = \tau_0$, the hydrodynamics turn-on time; $\tau_0$ is a particularly reasonable choice since we already only consider the medium density evolution for times $\tau>\tau_0$ in computing our pathlengths, \cref{eqn:effective_length}.  For a typical value of $\mu = 0.5~\mathrm{GeV}$ in central $\mathrm{Pb} + \mathrm{Pb}$ collisions, $\tau_0 \simeq \mu^{-1} \simeq 0.4~\mathrm{fm}$; however in central $\mathrm{p} + \mathrm{Pb}$ collisions $\mu^{-1} \simeq 0.25 ~\mathrm{fm} < \tau_0 = 0.4~\mathrm{fm}$, and so the distinction between the two options for $a$ might be important. In this report we have chosen to use $a = \tau_0$ throughout for simplicity. 

\begin{figure}[H]
    \centering
    \includegraphics[width=\linewidth]{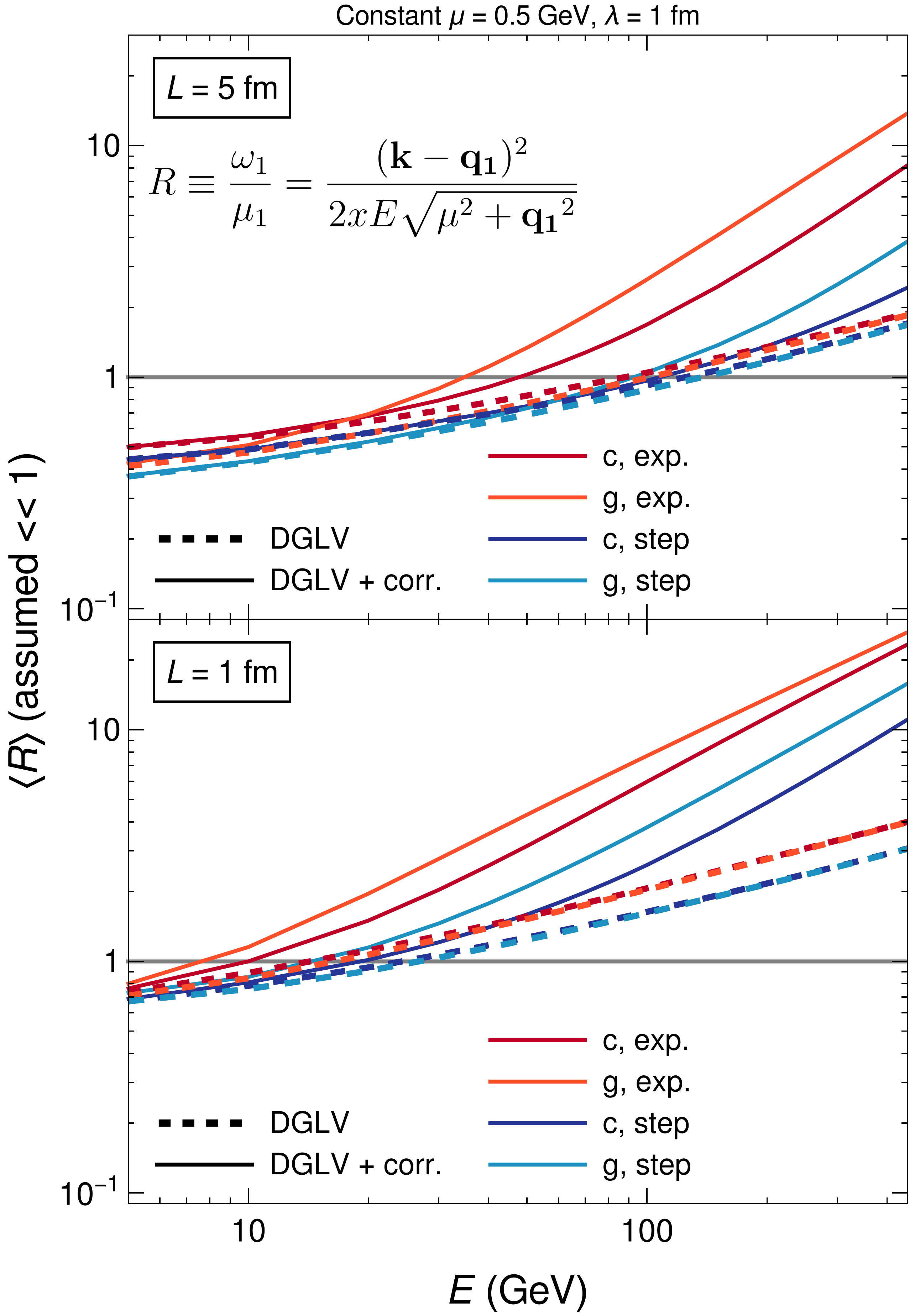}
    \caption{ 
   Plot of $\langle R\rangle \equiv \langle \omega_1/\mu_1 \rangle$ as a function of parent parton energy $E$.  $\langle R \rangle\ll1$ implies consistency with the large formation time assumption.  $\langle R \rangle$ is computed without (dashed) and with (solid) the short pathlength correction for charm quarks [gluons] with scattering centers distributed according to the exponential distribution (dark red [orange]) and truncated step function (dark blue [light blue]).  $L=5$ fm in the top pane and $L=1$ fm in the bottom pane.  All curves use constant $\lambda_g=1~\mathrm{fm}$ and $\mu=0.5~\mathrm{GeV}$.}
    \label{fig:largeformationtime2}
\end{figure}

\begin{figure}[H]
  \centering
  \includegraphics[width=\linewidth]{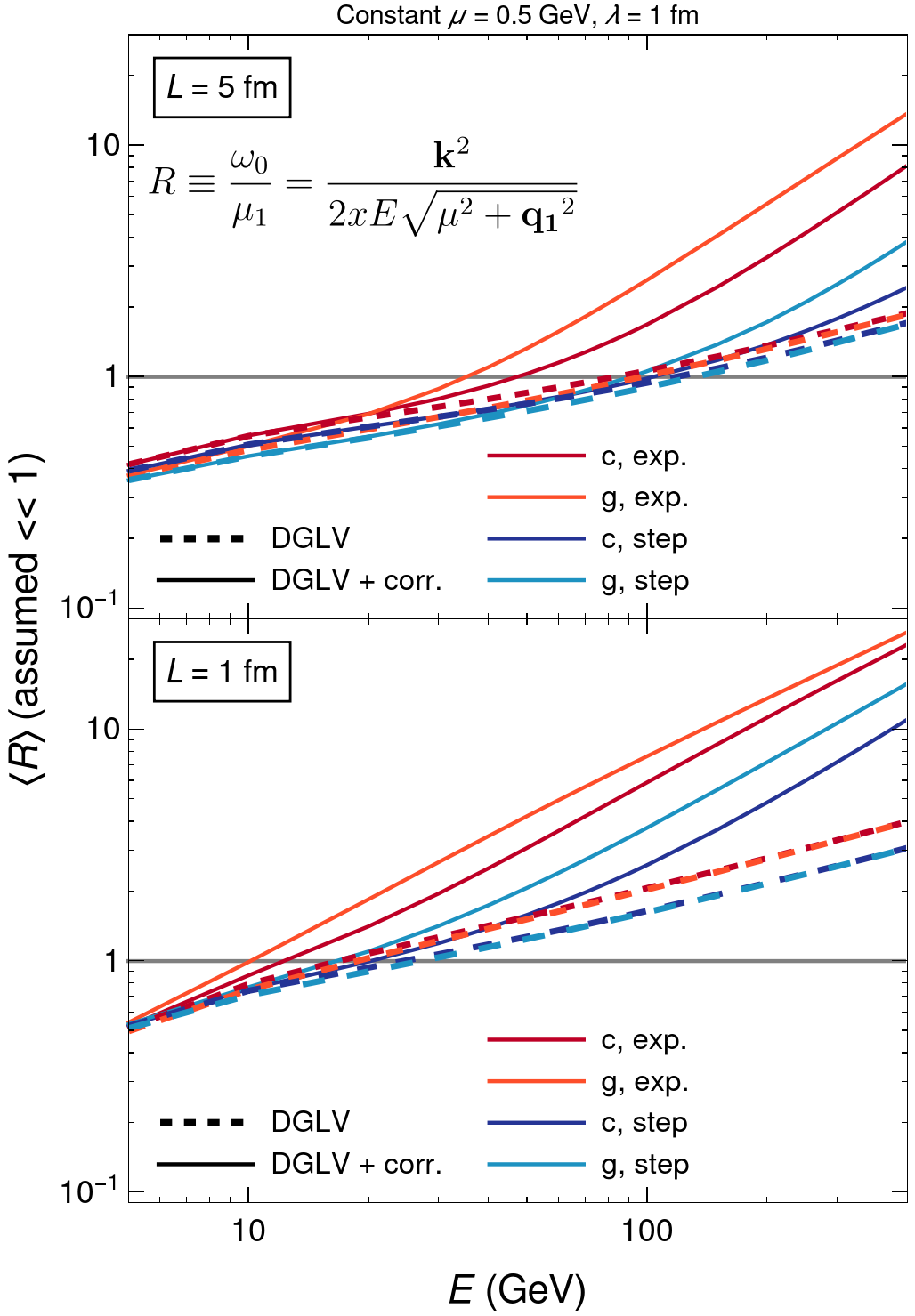}
    \caption{
   Plot of $\langle R\rangle \equiv \langle \omega_0/\mu_1 \rangle$ as a function of parent parton energy $E$.  $\langle R \rangle\ll1$ implies consistency with the large formation time assumption.  $\langle R \rangle$ is computed without (dashed) and with (solid) the short pathlength correction for charm quarks [gluons] with scattering centers distributed according to the exponential distribution (dark red [orange]) and truncated step function (dark blue [light blue]).  $L=5$ fm in the top pane and $L=1$ fm in the bottom pane.  All curves use constant $\lambda_g=1~\mathrm{fm}$ and $\mu=0.5~\mathrm{GeV}$.}
  \label{fig:largeformationtime3}
\end{figure}

\begin{figure}[H]
    \centering
    \includegraphics[width=\linewidth]{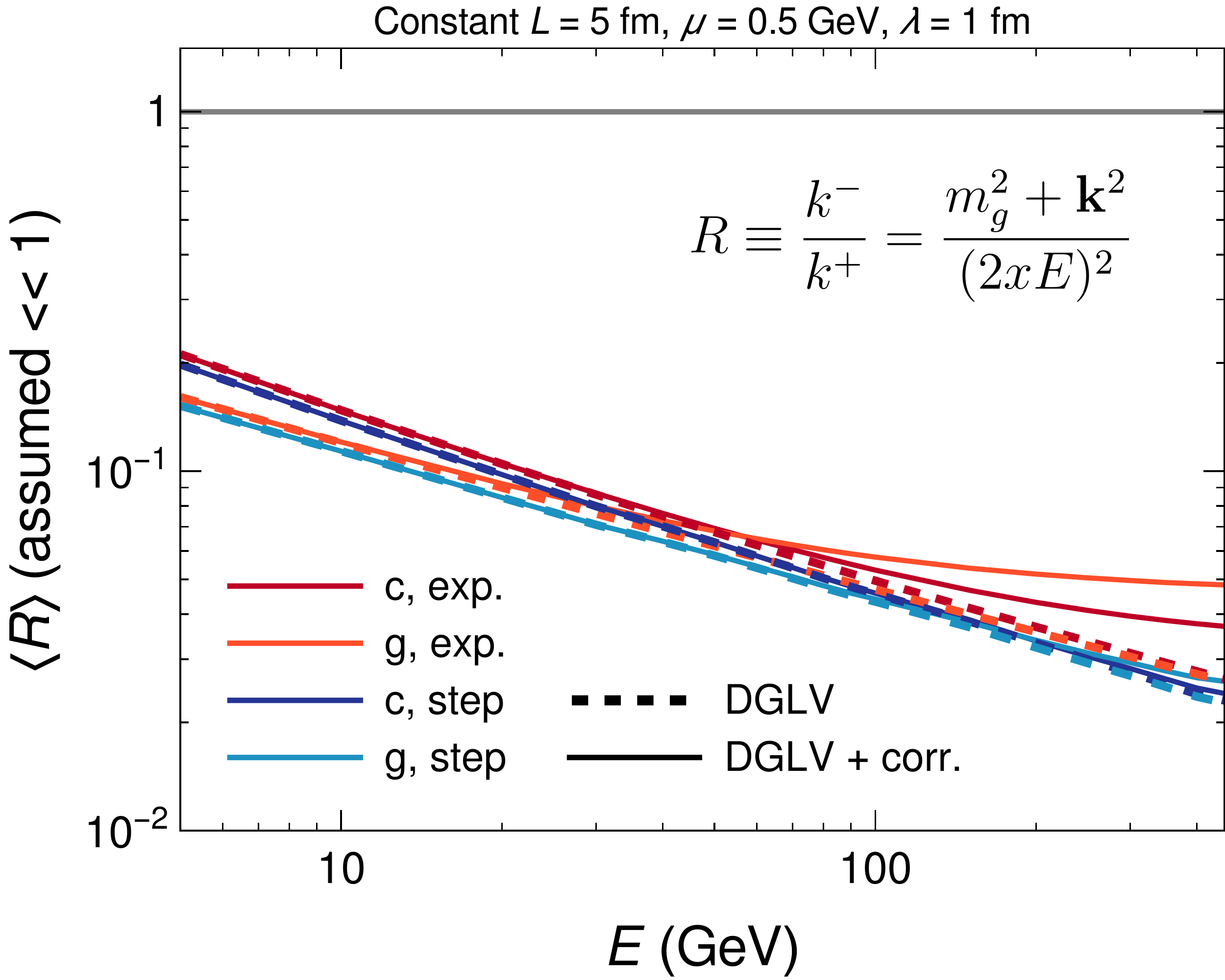}
    \caption{
   Plot of $\langle R\rangle \equiv \langle k^-/k^+ \rangle$ as a function of parent parton energy $E$.  $\langle R \rangle\ll1$ implies consistency with the collinear approximation.  $\langle R \rangle$ is computed without (dashed) and with (solid) the short pathlength correction for charm quarks [gluons] with scattering centers distributed according to the exponential distribution (dark red [orange]) and truncated step function (dark blue [light blue]).  $L=5$ fm in the top pane and $L=1$ fm in the bottom pane.  All curves use constant $L=5$ fm, $\lambda_g=1~\mathrm{fm}$, and $\mu=0.5~\mathrm{GeV}$.}
    \label{fig:collinear}
\end{figure}

\begin{figure}[H]
    \centering
    \includegraphics[width=\linewidth]{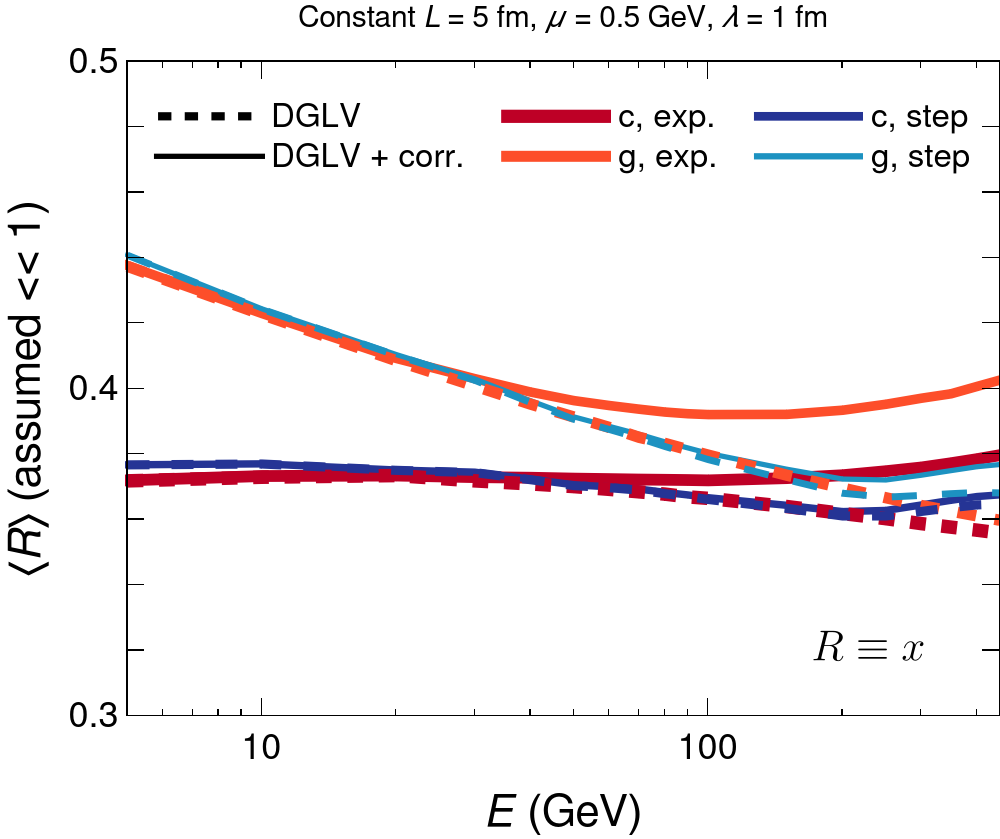}
    \caption{
   Plot of $\langle R\rangle \equiv \langle x \rangle$ as a function of parent parton energy $E$.  $\langle R \rangle\ll1$ implies consistency with the soft approximation.  $\langle R \rangle$ is computed without (dashed) and with (solid) the short pathlength correction for charm quarks [gluons] with scattering centers distributed according to the exponential distribution (dark red [orange]) and truncated step function (dark blue [light blue]).  $L=5$ fm in the top pane and $L=1$ fm in the bottom pane.  All curves use constant $L=5$ fm, $\lambda_g=1~\mathrm{fm}$, and $\mu=0.5~\mathrm{GeV}$.}
    \label{fig:soft}
\end{figure}

\begin{figure}[H]
  \centering
  \includegraphics[width=\linewidth]{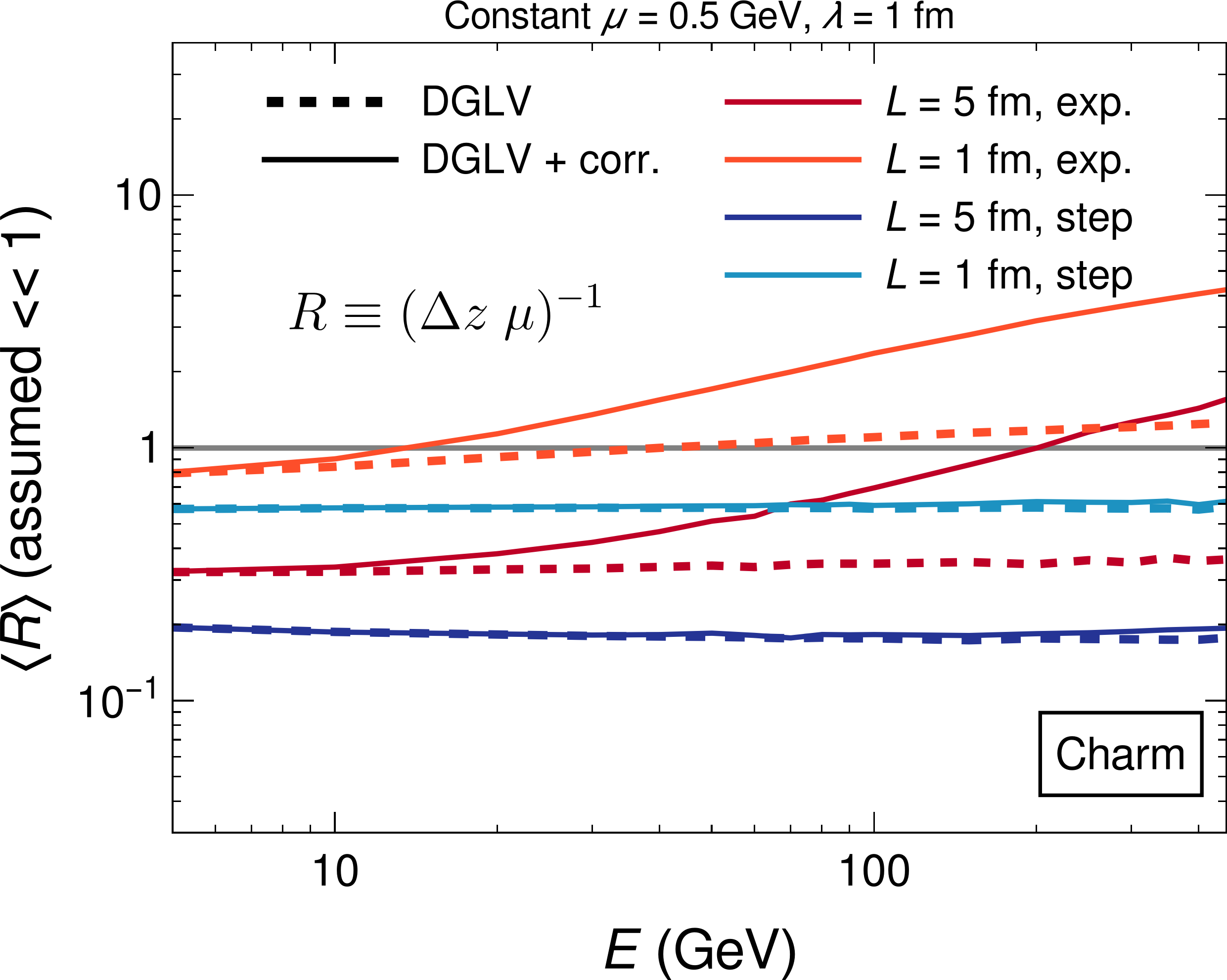}
    \caption{
   Plot of $\langle R\rangle \equiv \langle 1/\Delta z \, \mu \rangle$ as a function of parent par- ton energy $E$.  $\langle R \rangle\ll1$ implies consistency with the large pathlength assumption.  $\langle R \rangle$ is computed without (dashed) and with (solid) the short pathlength correction for charm quarks [gluons] with scattering centers distributed according to the exponential distribution (dark red [orange]) and truncated step function (dark blue [light blue]).  $L=5$ fm in the top pane and $L=1$ fm in the bottom pane.  All curves use constant $\lambda_g=1~\mathrm{fm}$ and $\mu=0.5~\mathrm{GeV}$.}
  \label{fig:largepathlengthassumption}
\end{figure}

\begin{figure}[H]
  \centering
  \includegraphics[width=\linewidth]{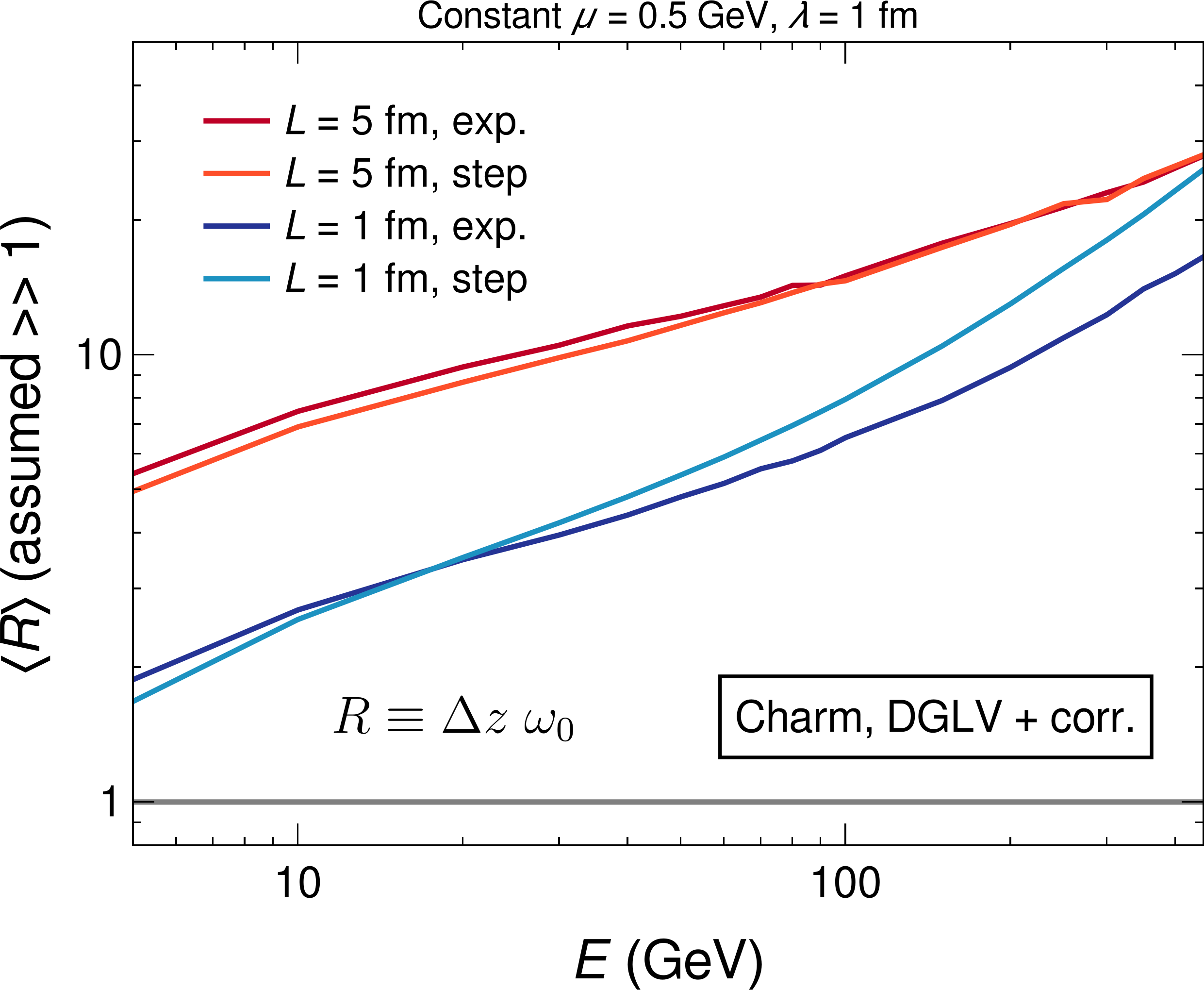}
    \caption{Plot of $\langle R\rangle \equiv \langle \Delta z \, \omega_0 \rangle$ as a function of parent parton energy $E$.  $\langle R \rangle\gg1$ implies consistency with the large formation time (with respect to scattering centers) assumption used in the short pathlength correction derivation \protect\cite{Kolbe:2015rvk}.  $\langle R \rangle$ is computed with the short pathlength correction for charm quarks for pathlengths of 5 fm [1 fm] with scattering centers distributed according to the exponential distribution (dark red [dark blue]) and truncated step function (orange [light blue]).  All curves use constant $\lambda_g=1~\mathrm{fm}$ and $\mu=0.5~\mathrm{GeV}$.}
  \label{fig:implicitlargepathlengthassumption}
\end{figure}

We now check all of the assumptions made in the computation of the DGLV radiative energy loss for: the corrected and uncorrected results; the exponential and truncated step scattering center distributions; charm quarks and gluons; and for large ($L = 5~\mathrm{fm}$) and small ($L = 1~\mathrm{fm}$) systems (\crefrange{fig:largeformationtime2}{fig:implicitlargepathlengthassumption}). All calculations use constant $\mu=0.5~\mathrm{GeV}$, and $\lambda_g = 1~\mathrm{fm}$ which are approximate averages in $\mathrm{A} + \mathrm{A}$ collisions, and standard benchmark choices \cite{Armesto:2011ht}. 

\sloppy\cref{fig:largeformationtime2,fig:largeformationtime3} show the expectation values of $\omega_1 / \mu_1 = (\mathbf{k} - \mathbf{q}_1)^2 / 2xE \sqrt{\mu^2 + \mathbf{q}_1^2}$ and $\omega_0 / \mu_1 = \mathbf{k}^2/2xE\sqrt{\mu^2+\mathbf{q}_1^2}$, respectively.  The large formation time assumption is equivalent to both ratios much less than one. For an exponential distribution of scattering centers with $L = 5~\mathrm{fm}$ (top panes of \cref{fig:largeformationtime2,fig:largeformationtime3}), the large formation time assumption: breaks down for $E \gtrsim 40~\mathrm{GeV}$ for the corrected result, and breaks down for $E \gtrsim 100$ GeV for the uncorrected result. For the truncated step distribution with $L = 5~\mathrm{fm}$, the large formation time assumption breaks down for both the corrected and uncorrected results for $E \gtrsim 100~\mathrm{GeV}$. We see in the plots the known numerical insensitivity to the distribution of scattering centers in the uncorrected DGLV radiative energy loss result.

For $L = 1~\mathrm{fm}$ (bottom panes of \cref{fig:largeformationtime2,fig:largeformationtime3}), the shape of all curves are approximately the same, but scaled by a factor $\sim1~\mathrm{fm} / 5~\mathrm{fm}$ in $E$. Thus the breakdown in the large formation time assumptions occurs roughly five times earlier in $E$ for the $L = 1$ fm pathlengths compared to the $L=5$ fm pathlengths. The reason for this simple approximate scaling is that all of the nontrivial dependence of $\Delta E / E$ on $E$ and $L$ in the distribution of emitted gluons \cref{eqn:full_dndx} comes from terms $\sim \omega_\alpha \Delta z$ where $\alpha \in \{0, 1, m\}$. Once integrated these terms become $\omega_\alpha L \sim L / E$; assuming $k$ and $q$ have negligible dependence on $E$. If the kinematic cutoffs on the $\mathbf{k}$, and $\mathbf{q}$ integrals are important then this scaling breaks down; any deviation from this simple scaling must be due to the effects of the cutoff. 

We found that finite kinematic bounds do not significantly affect the consistency of the assumptions. The above scaling argument holds for all expectation values $\langle R \rangle$ so long as $R$ does not depend on $\Delta z$. Thus, in order to keep the number of plots shown to a manageable number, most assumption consistency plots from now on will be shown for only $L=5~\mathrm{fm}$.

The collinear and soft assumptions are tested for consistency in \cref{fig:collinear,fig:soft}, respectively. We find that both of these assumptions are consistently satisfied for both the short pathlength corrected and uncorrected DGLV results, for both the exponential and the truncated step distributions of scattering centers.

\Cref{fig:soft} shows $\langle x \rangle$ as a function of parent parton energy $E$, where $\langle x \rangle \ll 1$ is assumed under the soft approximation. The expectation value $\langle x \rangle$ decreases monotonically in energy for the uncorrected result with an exponential distribution of scattering centers; all other expectation values appear to converge numerically to a constant nonzero value. 
For the DGLV result with an exponential distribution, one can calculate $\langle x \rangle$ analytically for asymptotically high energies. For asymptotically high energies we take $m_g \to 0$, $M \to  0$, $k_{\text{max}} \to \infty$, and $q_{\text{max}}\to \infty$. These simplifications allow the angular, $\mathbf{k}$, and $\mathbf{q}$ integrals to be done analytically, as described in \cite{Gyulassy:2000er, Djordjevic:2003zk}. Proceeding in this way we obtain the following asymptotic expression for $\langle x \rangle $:
\begin{align}
  \langle x \rangle^{\text{DGLV}}_{\text{exp.}} &=  \frac{1}{\log (\frac{4 E}{L \mu^2})} + \mathcal{O}\left( \frac{L \mu^2}{4 E} \right)\label{eqn:asymptotic_x_DGLV_exp}\\
  \implies \langle x \rangle &\to 0 \text{ as } E \to \infty.\nonumber
\end{align}

In a similar way we can derive the same result for the short pathlength correction with an exponential distribution, using the asymptotic result from \cite{Kolbe:2015rvk} (see \cref{eqn:asymptotics_correction_xint}) 
\begin{align}
  \langle x \rangle^{\text{corr.}}_{\text{exp.}} &= \frac{1}{2} \left[\frac{-\frac{1}{2}  + \log\left(\frac{2 E L}{2+L\mu}\right)}{-1+\log\left(\frac{2 E L}{2+L\mu}\right)}\right]\label{eqn:asymptotic_x_Short_exp}\\
  &\to \frac{1}{2} \text{ as } E\to \infty. \nonumber
\end{align}
Note that numerical investigation of $\langle x \rangle$ indicates that the convergence to the asymptotic values is slow for both the corrected and uncorrected energy loss.
For the truncated step distribution it is more difficult to perform asymptotic calculations, but numerical investigation shows that the uncorrected result with truncated step distribution converges to $\langle x \rangle \approx \frac{1}{3}$, and the corrected result with truncated step distribution converges to $\langle x \rangle \approx \frac{1}{2}$. 

\Cref{fig:largepathlengthassumption} tests the consistency of the large pathlength assumption with the DGLV result as a function of parent parton energy $E$ for charm quarks. For $L=1~\mathrm{fm}$, we see that the large pathlength assumption is not a good approximation for either the uncorrected or small pathlength corrected DGLV calculation for the exponential scattering center distribution; with $\langle 1 / \Delta z \, \mu \rangle\sim 0.6$, the large pathlength assumption is not a particularly good approximation for the truncated step distribution, either. Even for $L=5~\mathrm{fm}$, the large pathlength assumption breaks down for the short pathlength corrected energy loss when the exponential distribution of scattering centers is used. While not shown, results for gluons are essentially identical.  We see that, as expected, one must quantitatively determine the importance of the short pathlength correction terms to the radiative energy loss derivation for short pathlengths, $L\sim1/\mu$ for all values of $E$ and for all parton types.

We note that there was one final assumption implicitly made in the derivation of the short pathlength correction to the DGLV radiated gluon distribution, the short formation time (with respect to scattering centers) assumption, $\Delta z \: \omega_0 \gg 1$ \cite{Kolbe:2015rvk, Kolbe:2015suq}. This in conjunction with the large formation time assumption furnishes a separation of scales $\Delta z^{-1} \ll \omega_0 \ll \mu_1$. Alternatively one can view this as a large pathlength assumption, wherein we have replaced $\Delta z \: \mu \gg 1$ with $\Delta z \: \omega_0 \gg 1$; which is guaranteed to be a weaker assumption, according to the large formation time assumption.  \cref{fig:implicitlargepathlengthassumption} shows $\langle\Delta z \: \omega_0\rangle$ as a function of parent parton energy $E$ for charm quarks with the short pathlength corrected energy loss.  We find that for both large and small systems, and exponential and truncated step distributions this assumption holds self-consistently. While not shown, results for gluons are essentially identical.  The short formation time (with respect to scattering centers) assumption is intimately tied to the large formation time assumption, and so if future work aims to remove the large formation time assumption, then this assumption should also be removed. Note that we computed $\langle \Delta z \: \omega_0 \rangle$ (assumed $\gg 1$) instead of $\langle (\Delta z \: \omega_0)^{-1} \rangle$ (assumed $\ll 1$), due to numerical convergence issues with the latter.

\section{Suppression with Exponential and Step Distributions}
\label{sec:final_results}

We saw in the previous section that for $p_T\gtrsim40$ GeV the large formation time approximation breaks down for energy loss calculations using the exponential distribution of scattering centers (which bias the scattering to shorter pathlengths) while the large formation time approximation breaks down only for $p_T\gtrsim100$ GeV for an energy loss calculation using the truncated step distribution of scattering centers.  Additionally, we claimed that the exponential distribution of scattering centers and the truncated step distribution of scattering centers represent two extreme possibilities for what a more realistic distribution of scattering centers likely will be.  We are thus motivated to explore the sensitivity of our suppression predictions to the choice of distribution of scattering centers.  For pure DGLV energy loss without the short pathlength correction, one saw an insensitivity to this choice of scattering center distributions \cite{Armesto:2011ht}.  We will see that when including the short pathlength correction to the radiative energy loss, the heavy flavor observables are still relatively insensitive to the distribution of scattering centers.  However, we find that hadron observables that include a contribution from gluons are very sensitive to the choice of scattering center distribution, with the use of the truncated step function dramatically reducing the effect of the short pathlength correction to pion $R_{AA}(p_T)$.  This dramatic reduction in the effect of the short pathlength correction to the phenomenologically accessible pion suppression observable is expected from the dramatic reduction in the effect of the short pathlength correction to the average radiative energy loss, as seen in \cite{Kolbe:2015rvk}.

\begin{figure}[H]
  \centering
  \includegraphics[width=\linewidth]{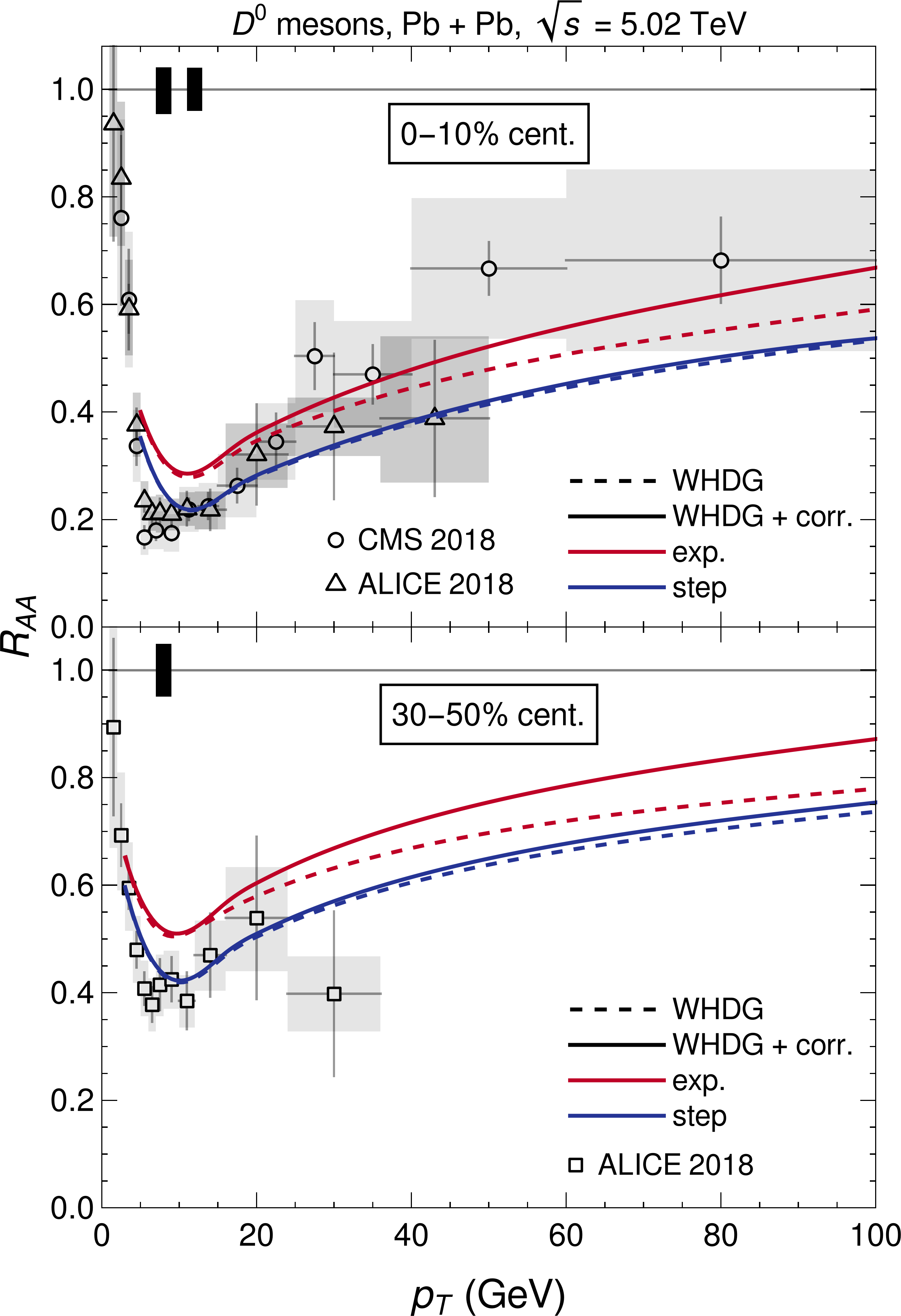}
  \caption{Plot of the $R_{AA}$ for $D$ mesons as a function of final transverse momentum $p_T$ in $\sqrt{s} = 5.02$ TeV $\mathrm{Pb}$+$\mathrm{Pb}$ central 0--10\% (top) and semi-central 30--50\% (bottom) collisions. Predictions with (solid) and without (dashed) the short pathlength correction to the radiative energy loss are shown using the exponential (red) and truncated step (blue) distributions for the scattering centers. Data are from ALICE \cite{ALICE:2018lyv} and CMS \cite{CMS:2017uoy}. The global normalization uncertainty on the number of binary collisions is indicated by solid boxes in the top left corner of the plot (left to right, top to bottom: CMS 0--10\%, ALICE 0--10\%, ALICE 30--50\%).}
  \label{fig:raa_D_mesons_with_step}
\end{figure}

\begin{figure}[H]
  \centering
  \includegraphics[width=\linewidth]{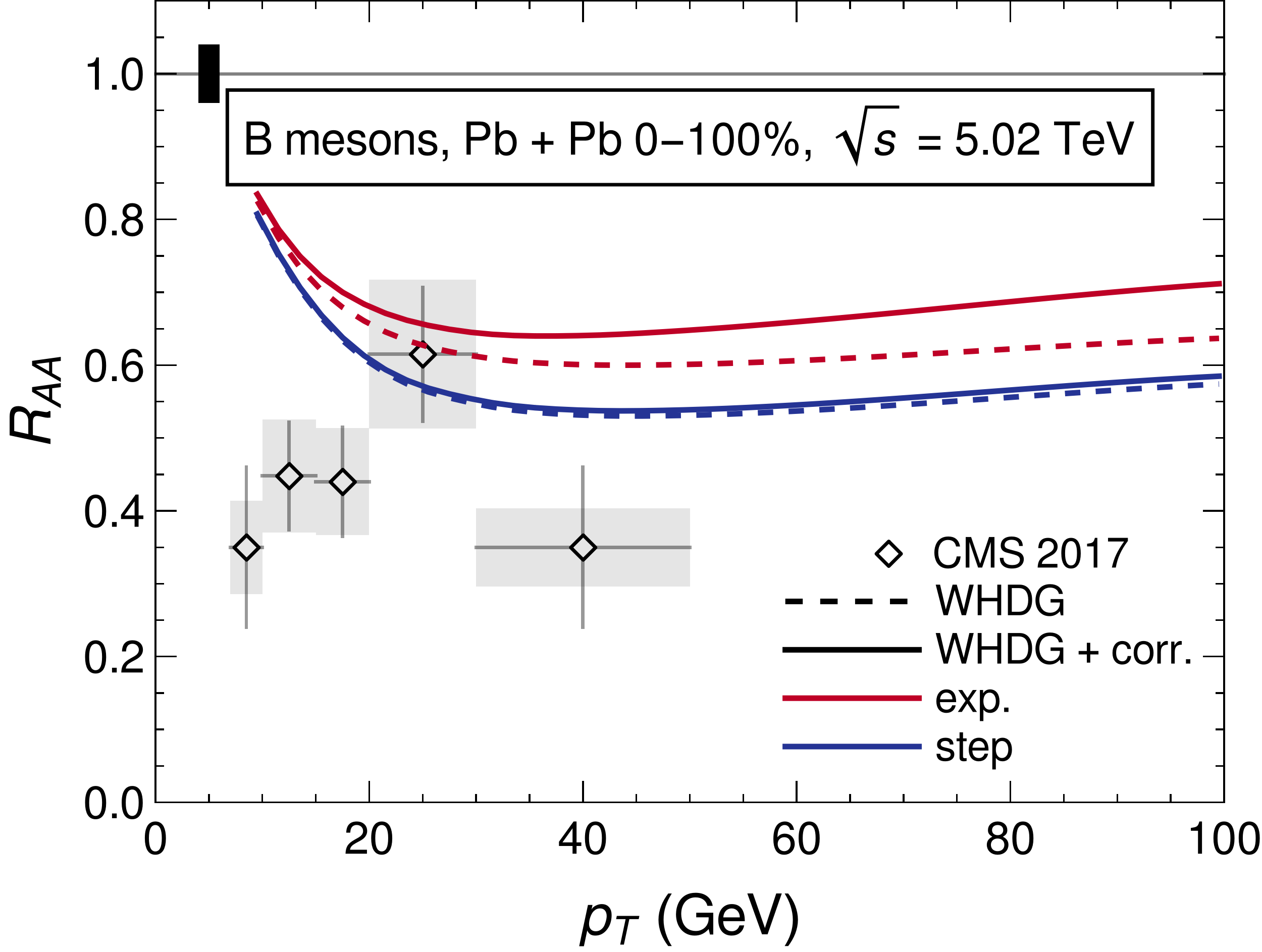}
  \caption{ The nuclear modification factor $R_{AA}$ as a function of final transverse momentum $p_T$ is calculated in $\mathrm{Pb} + \mathrm{Pb}$ collisions at $\sqrt{s}=5.02$ TeV for $B$ mesons. Predictions with (solid) and without (dashed) the short pathlength correction to the radiative energy loss are shown using the exponential (red) and truncated step (blue) distributions for the scattering centers. Data are from CMS \cite{CMS:2017uoy}. The experimental global normalization uncertainty on the number of binary collisions is indicated by the solid box in the top left corner of the plot.}
  \label{fig:raa_B_mesons_with_step}
\end{figure}

\Cref{fig:raa_D_mesons_with_step} shows $R_{AA}(p_T)$ for $D$ mesons in central 0--10\% and semi-central 30--50\% $\mathrm{Pb}+\mathrm{Pb}$ collisions at $\sqrt{s}=5.02$ TeV for both the exponential and truncated step distributions of scattering centers. \Cref{fig:raa_B_mesons_with_step} shows $R_{AA}(p_T)$ for $B$ mesons in central 0--10\% $\mathrm{Pb}+\mathrm{Pb}$ collisions at $\sqrt{s}=5.02$ TeV for both the exponential and truncated step distributions of scattering centers.  For both heavy meson cases the truncated step distribution decreases the $R_{A A}$ by up to 10\% (20\%) for the original (corrected) WHDG results. For both $D$ and $B$ mesons with a truncated step distribution, the short pathlength correction is negligible for $p_T \lesssim 100$. As seen before, with an exponential distribution of scattering centers the effect of the short pathlength correction is $\lesssim 10\%$, which is small compared to other theoretical uncertainties in the model (e.g.\ higher orders in $\alpha_s$, treatment of the early times, etc.). Agreement with data is good for all predictions (corrected/uncorrected, and exp./trunc. step distributions), except for $p_T \lesssim 10$ where bulk effects may be important and the eikonal approximation is likely breaking down.

\begin{figure}[H]
  \centering
  \includegraphics[width=\linewidth]{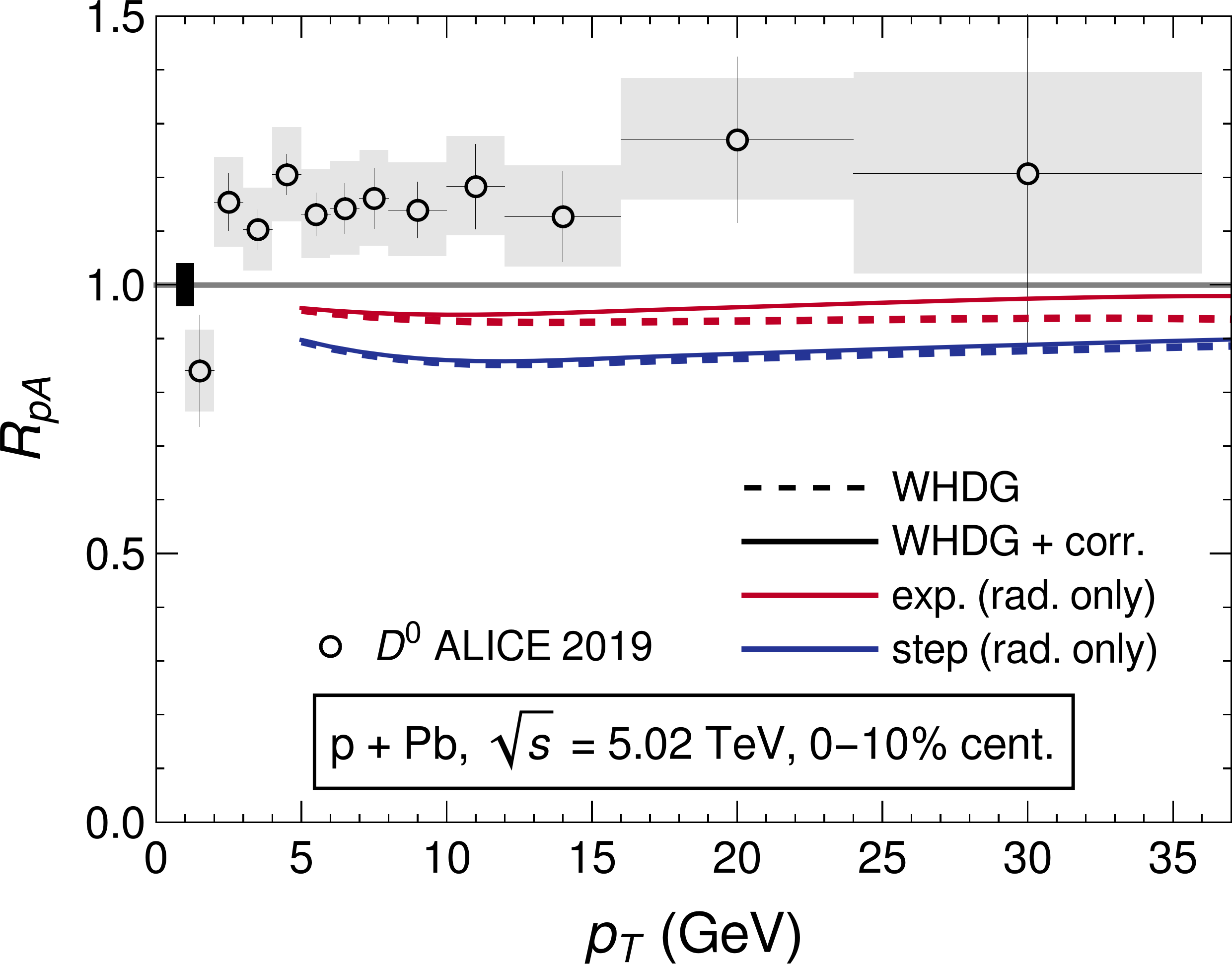}
    \caption{The nuclear modification factor $R_{pA}$ for $D$ mesons as a function of final transverse momentum $p_T$ in 0--10\% central $\mathrm{p} + \mathrm{Pb}$ collisions at $\sqrt{s}=5.02$ TeV.  Only radiative energy loss is included; predictions with (solid) and without (dashed) the short pathlength correction are shown using the exponential (red) and truncated step (blue) distributions for the scattering centers.  Data are from ALICE \cite{ALICE:2019fhe}. The experimental global normalization uncertainty on the number of binary collisions is indicated by the solid box in the top left corner of the plot.}
  \label{fig:raa_D_mesons_pA_with_step}
\end{figure}

The radiative-only nuclear modification factor $R_{pA}$ for $D^0$ mesons in 0--10\% most central $\mathrm{p}+\mathrm{A}$ collisions is shown in \cref{fig:raa_D_mesons_pA_with_step}, with and without the short pathlength correction, and with both an exponential and a truncated step distribution of scattering centers. We only show the radiative-only $R_{p A}$ since in \cref{sec:results} we determined that the elastic contribution was erroneously large in $\mathrm{p}+\mathrm{A}$ collisions due to the inapplicability of the average elastic energy loss in previous WHDG calculations in small collision systems. The $R_{pA}$ for the exponential distribution without the correction and the truncated step distribution with and without the correction all agree with each other to within $5\%$ and predict mild suppression of $R_{pA} \approx 0.9$ for all $p_T$. The corrected $R_{pA}$ with an exponential distribution predicts mild suppression at low $p_T \lesssim 20 ~\mathrm{GeV}$ and consistency with unity at moderate $p_T \gtrsim 20 ~\mathrm{GeV}$. Measured $R_{p A}$ is shown for $D^0$ mesons produced in 0--5\% most central $\mathrm{p} + \mathrm{Pb}$ collisions from ALICE \cite{ALICE:2019fhe}. Data predicts mild enhancement for all $p_T$, not inconsistent with our results shown here.

\begin{figure}[H]
  \centering
  \includegraphics[width=\linewidth]{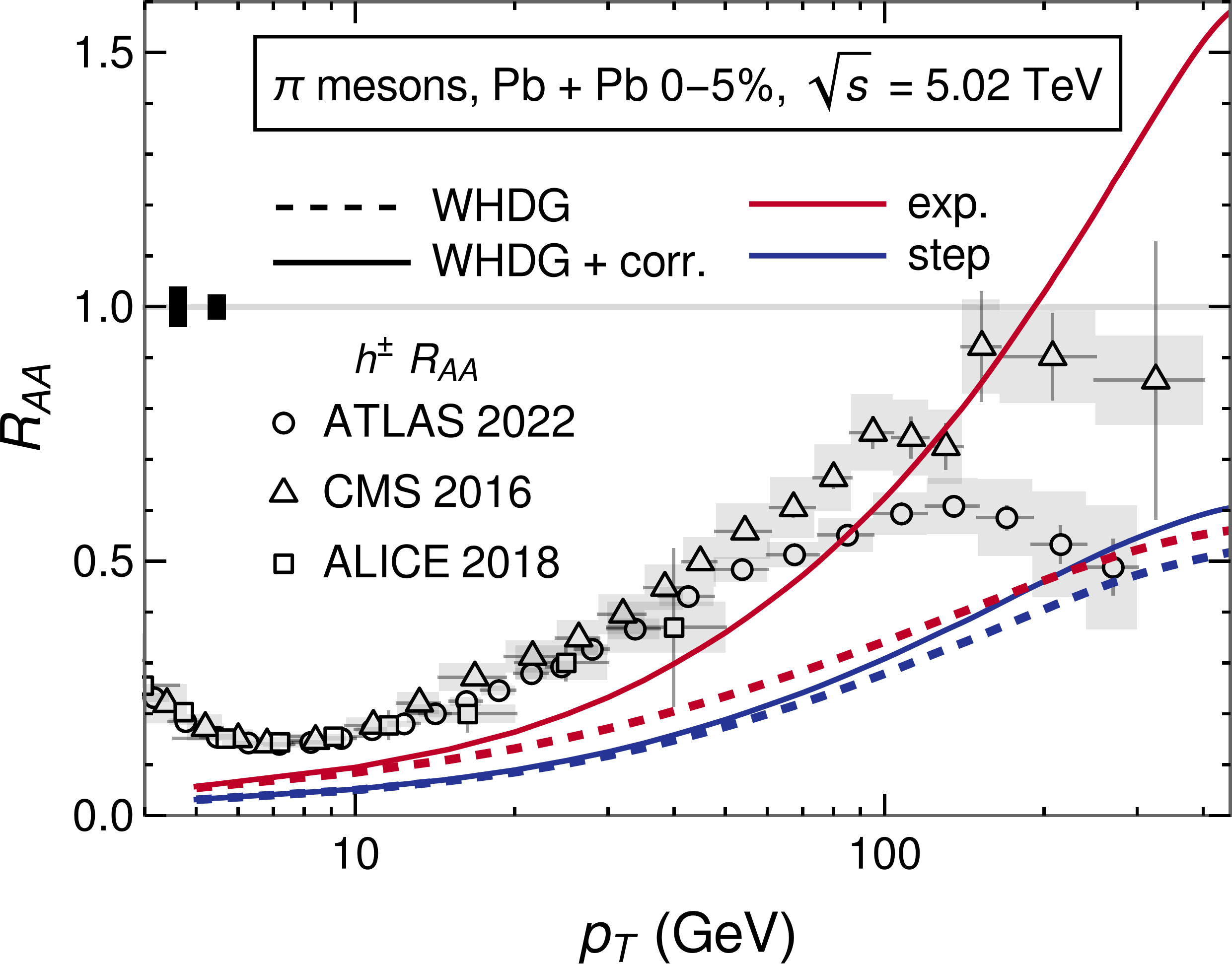}
  \caption{Plot of the  $R_{AA}$ for $\pi$ mesons produced in 0--5\% most central $\mathrm{Pb}+\mathrm{Pb}$ collisions at $\sqrt{s} = 5.02$ TeV, as a function of the final transverse momentum $p_T$. Predictions with (solid) and without (dashed) the short pathlength correction to the radiative energy loss are shown using the exponential (red) and truncated step (blue) distributions for the scattering centers.  Data from ATLAS \cite{ATLAS:2022kqu}, CMS \cite{CMS:2016xef}, and ALICE \cite{Sekihata:2018lwz}. The experimental global normalization uncertainty on the number of binary collisions is indicated by the solid boxes in the center left of the plot (left to right: CMS, ALICE). Note that in the ATLAS data the normalization uncertainty is included in the systematic uncertainty.}
  \label{fig:raa_pions_with_step}
\end{figure}

\Cref{fig:raa_pions_with_step} shows $R_{AA}(p_T)$ for pions in 0--5\% central $\mathrm{Pb}+\mathrm{Pb}$ collisions at $\sqrt{s}=5.02$ TeV.  The convolved collisional and radiative energy loss model predictions both include and exclude the short pathlength correction to radiative energy loss, and use either the exponential or the truncated step distribution for the scattering centers.  The predictions are compared to data from ATLAS \cite{ATLAS:2022kqu}, CMS \cite{CMS:2016xef}, and ALICE \cite{Sekihata:2018lwz}.  While the predictions excluding the short pathlength correction are insensitive to the particular scattering center distribution chosen, one can see the tremendous sensitivity of the predictions to the scattering center distribution when the short pathlength correction to the radiative energy loss is included; when the truncated step distribution is used, the effect on $R_{AA}(p_T)$ of the short pathlength correction to the DGLV radiative energy loss is dramatically reduced.  

\begin{figure}[H]
  \centering
  \includegraphics[width=\linewidth]{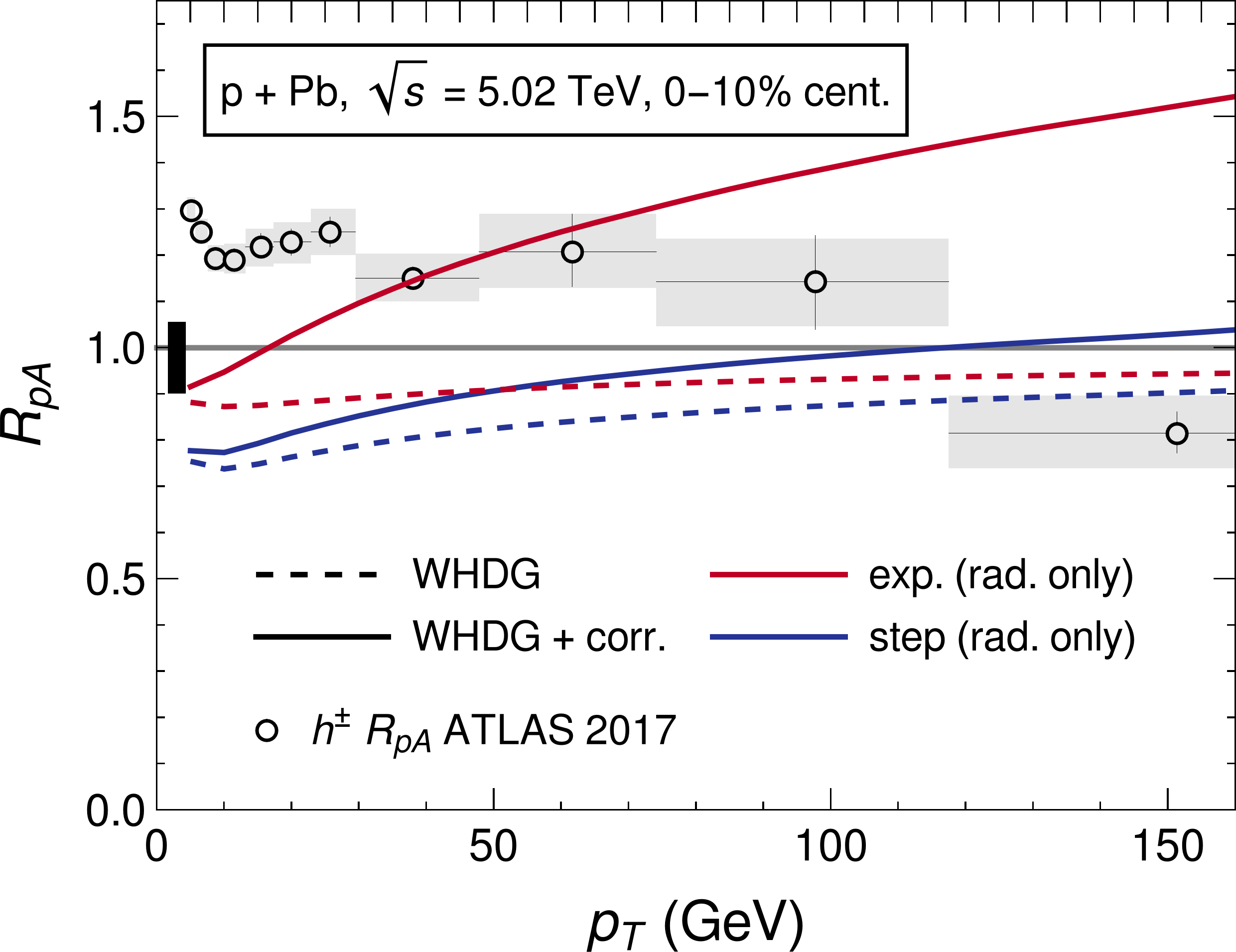}
    \caption{ The nuclear modification factor $R_{p A}$ for $\pi$ mesons as a function of final transverse momentum $p_T$ in 0--10\% $\mathrm{p}+\mathrm{Pb}$ collisions at $\sqrt{s}=5.02$ TeV.   Only radiative energy loss is included; predictions with (solid) and without (dashed) the short pathlength correction are shown using the exponential (red) and truncated step (blue) distributions for the scattering centers.  Data for charged hadrons are from ATLAS \protect\cite{Balek:2017man}. The experimental global normalization uncertainty on the number of binary collisions is indicated by the solid box in the top left corner of the plot.
    }
  \label{fig:raa_pi_mesons_pA_with_step}
\end{figure}

\Cref{fig:raa_pi_mesons_pA_with_step} shows $R_{p A}(p_T)$ for $\pi$ mesons and charged hadrons in 0--10\% central $\mathrm{p}+\mathrm{Pb}$ collisions at $\sqrt{s}=5.02$ TeV.  For our theoretical predictions of pion suppression we only include radiative energy loss, as the average elastic energy loss of the WHDG model is inappropriate to use here.  We show predictions with (solid) and without (dashed) the short pathlength correction to the radiative energy loss, and we show predictions when using either the exponential (red) or truncated step (blue) distribution of scattering centers.  Charged hadron suppression data is from ATLAS \protect\cite{Balek:2017man}.

The difference between the $R^\pi_{pA}(p_T)$ predictions from the two scattering center distributions is small when excluding the short pathlength correction.  We see again that the effect on nuclear modification factor from the short pathlength correction to the radiative energy loss is large when using the exponential distribution of scattering centers and relatively small when using the truncated step distribution of scattering centers.  The prediction of enhancement by the corrected $R_{p A}$ with an exponential distribution is qualitatively similar to the observed enhancement for moderate momenta $p_T \lesssim 60~\mathrm{GeV}$. If we are provocative, we may thus suggest that the experimentally measured excess in $R_{pA}(p_T)$ is actually due to final state effects rather than initial state or normalization effects.

\section{Discussion}
\label{sec:discussion}

The primary goal of this work, as outlined in \crefrange{sec:energy_loss}{sec:final_results}, is to implement the WHDG convolved energy loss model \cite{Wicks:2005gt, Horowitz:2011gd} with the novel inclusion of the short pathlength correction to the radiative energy loss \cite{Kolbe:2015rvk}.

Due to the complexity of this model, there are many points at which one must decide on the level of approximation to proceed with. In line with the motivation of this work, we have always chosen to maintain consistency with previous work such as WHDG and DGLV. This has occasionally led us to overlook a more physically reasonable prescription (in our view) for a component of the energy loss model. One such instance concerns the treatment of the realistic collision geometry (see \cref{sec:geometry}) and how it is mapped to the brick geometry. We will now present a derivation of a more realistic effective pathlength and effective temperature, which could be implemented in future work.

An integral part of radiative energy loss is the distribution of scattering centers, $\bar{\rho}(\Delta z)$, normalized to a single hard scatter at first order in opacity (see \cref{eqn:density_scattering_centers}). The quantity $\bar{\rho}(\Delta z)$ is a model for the shape of the plasma, as the parton propagates through it. In theory, it is possible to improve the realism of the model by integrating through the plasma, in some sense replacing $\bar{\rho}(\Delta z)$ with a realistic plasma density $\rho(\vec{x}, \tau)$ and correspondingly $T$ with a realistic plasma temperature $T(\vec{x}, \tau)$. In this case the pathlength $L$ no longer needs to be specified \emph{a priori}. This approach would allow for a more accurate simulation of the hard partons propagation through the plasma, and is implemented similarly in the CUJET model \cite{Buzzatti:2011vt, Xu:2014ica}. Unfortunately implementing this approach is exceedingly computationally expensive, even for the original DGLV result, and would present significant computational challenges for the short pathlength corrected results. 

To account for a realistic collision geometry, we instead need \emph{effective} temperatures, densities, and lengths as inputs to the simple brick models for elastic and radiative energy loss. Essentially, we establish a brick with characteristic $\{L_{\text{eff}}, T_{\text{eff}}, \bar{\rho}\}$ for each parton that propagates through the plasma. %
The relationship between the plasma density $\rho(\mathbf{x}, \tau)$ and the distribution of scattering centers $\bar{\rho}(\Delta z)$, is a separation of the shape of the plasma density from its magnitude. Schematically, we can write this separation as
\begin{align}
  \frac{\mathrm{d}N^g}{\mathrm{d}x} =& \int \mathrm{d} \Delta z \; \rho(\mathbf{x}_i + \boldsymbol{\hat{\phi}}\Delta z, \Delta z) \left.\frac{\mathrm{d} N^{g}}{\mathrm{d}\Delta z \, \mathrm{d} x}\right|_{\mu = \mu(z)}
  \label{eqn:CUJET_dNdX_schematic}\\
  \approx& \int \mathrm{d} \Delta z' \; \rho(\mathbf{x}_i + \boldsymbol{\hat{\phi}}\Delta z', \Delta z') \nonumber\\
  &\times \int \mathrm{d} \Delta z \; \bar{\rho}(\Delta z) \left.\frac{\mathrm{d} N^{g}}{\mathrm{d}\Delta z \, \mathrm{d}x} \right|_{\mu = \mu_{\text{eff}}},
  \label{eqn:DGLV_dNdX_schematic}
\end{align}
where $\int \mathrm{d} \Delta z \; \bar{\rho}(\Delta z) = 1$. In the above $\mathbf{x}_i$ is the hard parton production point, $\boldsymbol{\hat{\phi}}$ is the direction of propagation, and $\rho$ is the density of the QGP. \Cref{eqn:CUJET_dNdX_schematic} is equivalent to what is done in CUJET \cite{Buzzatti:2011vt, Xu:2014ica} while \cref{eqn:DGLV_dNdX_schematic} is the approximation made in GLV \cite{Gyulassy:2000er}, DGLV \cite{Djordjevic:2003zk}, WHDG \cite{Wicks:2005gt, Horowitz:2011gd} and the short pathlength correction to DGLV \cite{Kolbe:2015rvk}. Note that the step from \cref{eqn:CUJET_dNdX_schematic} to \cref{eqn:DGLV_dNdX_schematic} is exact for a brick of plasma. 

The power of the approximation made in \cref{eqn:DGLV_dNdX_schematic} lies in separating the dependence of the path taken by the parton through the plasma from the rest of the energy loss calculation. This approach allows us to prescribe $\bar{\rho}(\Delta z)$, for instance exponential decay or truncated step, and perform the $\Delta z$ integral analytically. 
A more realistic approach to this, but less numerically intensive than integrating through the realistic plasma, would be to fit a trial $\bar{\rho}(\Delta z | \mathbf{x}_i, \phi)$ to the realistic plasma density $\rho(\mathbf{x}_i, \phi)$ for each path taken by a parton through the plasma.

The magnitude of the density is definitionally related to the opacity $\bar{n}$ via \cite{Gyulassy:2000er}
\begin{align}
  \bar{n} \equiv \frac{L_{\text{eff}}}{\lambda_{\text{eff}}} \equiv & \int d z \int d^2 \mathbf{q} \frac{d \sigma_{gg}(z)}{d^2 \mathbf{q}} \rho(\mathbf{x}_i + z \boldsymbol{\hat{\phi}}, \tau=z)\label{eqn:opacity_definition}\\
    \approx& ~ \sigma_{gg}^{\text{eff}} \int d z \;  \rho(\mathbf{x}_i + z \boldsymbol{\hat{\phi}}, \tau=z)
  \label{eqn:opacity_approximation}
\end{align}
where $\rho$ is the density from \cref{eqn:color_weighted_density}, and we have used \cref{eqn:opacity_approximation} to define the effective length $L_{\text{eff}}$ and effective gluon mean free path $\lambda_{\text{eff}}$. \Cref{eqn:mean_free_path,eqn:opacity_approximation} yield
\begin{align}
    \frac{L_{\text{eff}}}{\lambda_{\text{eff}}} =& \left(\lambda_{\text{eff}} \; \rho_{\text{eff}}\right)^{-1} \int \mathrm{d} z \; \rho(\mathbf{x}_i + z \boldsymbol{\hat{\phi}}, \tau=z)\\
    \implies L_{\text{eff}} =& \frac{1}{\rho_{\text{eff}}} \int \mathrm{d} z \; \rho(\mathbf{x}_i + z \boldsymbol{\hat{\phi}}, \tau=z),
  \label{eqn:effective_length2}
\end{align}
where we have the freedom to prescribe $\rho_{\text{eff}}$. Note that \cref{eqn:effective_length2} differs from \cref{eqn:effective_length} in the fact that the density is evaluated at $\tau = z$, which follows from the definition of the opacity in \cref{eqn:opacity_definition}.

Breaking apart the opacity $\bar{n}$ as $\bar{n} = L_{\text{eff}} / \lambda_{\text{eff}}$ serves to:
\begin{itemize}
  \item make contact with the effective pathlength prescription in WHDG \cite{Wicks:2005gt}, with a more rigorous derivation;
  \item obtain a prescription for the effective density $\rho_{\text{eff}}$ which can then be used to calculate other thermodynamic quantities in \cref{eqn:thermodynamic_quantities}, most importantly the Debye mass $\mu$;
  \item and obtain a length scale $L_{\text{eff}}$ in the problem, which is important for prescribing the distribution of scattering centers $\bar{\rho}(\Delta z)$.
\end{itemize}

In this manuscript, we have followed WHDG in using \cref{eqn:effective_density} as the effective density, which prescribes a single density to the entirety of the plasma. A more natural approach can be motivated by considering the step form \cref{eqn:CUJET_dNdX_schematic} to \cref{eqn:DGLV_dNdX_schematic}. In this step, we have approximated $\mu(z) \approx \mu_{\text{eff}}$, which leads to a natural definition for $\rho_{\text{eff}}$ as the average temperature along the path through the plasma, weighted by the plasma density.
\begin{equation}
  \rho_{\text{eff}} = \langle \rho \rangle(\mathbf{x}_i, \phi) = \frac{\int \mathrm{d} \Delta z \; \rho^2(\mathbf{x}_i + \boldsymbol{\hat{\phi}} \Delta z, \Delta z)}{\int \mathrm{d} \Delta z \; \rho(\mathbf{x}_i + \boldsymbol{\hat{\phi}} \Delta z, \Delta z)}.
  \label{eqn:effective_density2}
\end{equation}
This means both $L_{\text{eff}}$ and $\rho_{\text{eff}}$ depend on the specific path that the parton takes. This dramatically increases the numerical complexity as we must now evaluate the energy loss distribution for a distribution in $(L_{\text{eff}}, \rho_{\text{eff}})$. Note that Bjorken expansion is naturally taken into account with this prescription for the effective density $\rho_{\text{eff}}$ and so we do not need to use the approximation in \cref{eqn:bjorken_expansion}.

This approach could be implemented in future work, offering the advantage of a more realistic collision geometry compared to the implementation in this work and WHDG \cite{Wicks:2005gt}; while still being significantly less computationally expensive than integrating through the realistic plasma as in CUJET \cite{Xu:2014ica}.

\section{\label{sec:conclusion} Conclusions}

In this article we presented the first predictions for the suppression of leading high-$p_T$ hadrons from an energy loss model with explicit short pathlength corrections to the radiative energy loss.  We included collisional energy loss in the model, as well as averages over realistic production spectra for light and heavy flavor partons that propagate through a realistic QGP medium geometry generated by second order viscous hydrodynamics.  Thus our calculations here are, to a very good approximation, those of the WHDG energy loss model \cite{Wicks:2008zz}, but with short pathlength corrections \cite{Kolbe:2015rvk} to the DGLV opacity expansion \cite{Gyulassy:2000er,Djordjevic:2003zk}.  Predictions were presented for central and semi-central $\mathrm{Pb}+\mathrm{Pb}$ collisions and central $\mathrm{p}+\mathrm{Pb}$ collisions and compared to data from the LHC.  

We saw that the inclusion of the short pathlength correction to the radiative energy loss led to a \emph{reduction} of the suppression of leading hadrons.  This reduction is well understood as a result of the short pathlength correction enhancing the effect of the destructive LPM interference between the zeroth order in opacity DGLAP-like production radiation and the radiation induced by the subsequent collisions of the leading parton with the medium quanta.  The reduction in the suppression also increases as a function of $p_T$, which is a result of the different asymptotic energy scalings of DGLV energy loss ($\Delta E\sim\log E$) compared to the short pathlength correction ($\Delta E\sim E$).  

For heavy flavor observables, the inclusion of the short pathlength corrections leads to only a modest $\sim10$\% enhancement of $R_{AA}(p_T)$ in $\mathrm{Pb}+\mathrm{Pb}$ and $\mathrm{p}+\mathrm{Pb}$ collisions.  Even though the relative $R_{pA}$ enhancement of $\sim10\%$ is similar to that of $R_{AA}$, as one can see in \cref{fig:deltaE_vs_energy} the influence of the short pathlength correction on the energy loss is significantly larger for shorter pathlengths and therefore also in the smaller collision system.  The reason that the $R_{pA}$ and $R_{AA}$ have similar relative enhancements is due to the scaling $R_{AA}\sim(1-\epsilon)^{n-1}$, where $\epsilon\equiv\Delta E/E$ is the fractional energy lost by the leading parton.  One can determine that the effective short pathlength correction, averaged over the Poisson convolution and geometry, is about 100\% stronger in $\mathrm{p}+\mathrm{Pb}$ compared to central $\mathrm{Pb}+\mathrm{Pb}$.  (Note that the Poisson convolution, with its large probability of no interaction or energy loss for short pathlengths, is crucial for in fact \emph{reducing} the enormous short pathlength correction influence seen in \cref{fig:deltaE_vs_energy} in the energy loss model.)  Thus, since $R_{pA}\sim1$ and $n\sim6$, even though the short pathlength correction is about 100\% stronger in $\mathrm{p}+\mathrm{Pb}$, the influence on $R_{pA}$ is similar to that in $R_{AA}$.

When we assume that the distribution of scattering centers that stimulate the emission of gluon radiation from high-$p_T$ partons is given by an exponential, which biases the leading partons to scatter at shorter distances, the short pathlength correction to $R_{AA}^\pi(p_T)$ grows dramatically with $p_T$.  This very fast growth in $p_T$ is due to the very large short pathlength correction to the gluonic radiative energy loss: the short pathlength correction to radiative energy loss breaks color triviality, and the correction to gluonic radiative energy loss is about ten times that of quark radiative energy loss (instead of the approximately factor of two that one would expect from color triviality) \cite{Kolbe:2015rvk}.  When a truncated step function is used as the distribution of scattering centers that stimulate the emission of gluon radiation, the short pathlength correction to $R^\pi_{AA}(p_T)$ becomes a much more modest $\sim10$\%.  It is interesting, but perhaps not totally surprising, that the short pathlength correction to the energy loss introduces an enhanced sensitivity to the precise distribution of scattering centers used in the energy loss model.  In either case of distributions of scattering centers, the more rapid growth in the short pathlength correction as a function of $p_T$ than that of the uncorrected DGLV energy loss \cite{Horowitz:2011gd} suggests that at least part of the faster-than-expected growth of the measured $R^\pi_{AA}(p_T)$ as a function of $p_T$ may be due to the influence of the short distance corrections to the energy loss of hard partons; cf.\ the reduction of suppression due to running coupling \cite{Buzzatti:2012dy}. 

We found that the average collisional energy loss, with fluctuations of this energy loss given by a Gaussian whose width is dictated by the fluctuation-dissipation theorem, of the WHDG energy loss model \cite{Wicks:2008zz} is inappropriate for small colliding systems.  Considering radiative energy loss only, $R_{pA}(p_T)$ for heavy flavor hadrons was again only modestly, $\lesssim10$\%, affected by the short pathlength correction to energy loss.  $R_{pA}^\pi(p_T)$ was significantly affected, $\sim 50$\%, by the short pathlength correction for an exponential distribution of scattering centers, but more modestly so, $\sim 10$\%, for a truncated step distribution of scattering centers.  For both distributions of scattering centers, the short pathlength corrected pion nuclear modification factor sees a tantalizing enhancement above 1, similar to data \cite{Balek:2017man,ALICE:2018lyv}.  One may then provocatively suggest that the experimentally measured enhancement of $R_{pA}(p_T)>1$ may be due---at least in part---to final state effects.  

We also investigated the self-consistency of the approximations used in the derivations of DGLV and short pathlength correction to DGLV single inclusive radiative gluon emission kernels for the phenomenologically relevant physical situations of RHIC and LHC.  We constructed dimensionless quantities that represented the approximations and checked whether, when averaged with a weight given by the strength of the energy loss kernel determined by DGLV or the short pathlength corrected DGLV, those quantities were small (or large) as required by the approximations that went into deriving those same DGLV and short pathlength corrected DGLV single inclusive radiated gluon spectrum kernels.  

We found that, when weighted by the energy loss kernel, the soft and collinear approximations were self-consistently satisfied when computed with phenomenologically relevant parameters.  
We further found that both the original DGLV derivation and the DGLV derivation with the inclusion of the short pathlength correction are \emph{not} consistent with the large formation time approximation for modest $\mathcal O(10\text{--}100 \, \mathrm{GeV})$ energies and pathlengths $\mathcal O(1\text{--}5 \, \mathrm{fm})$, independent of the choice of distribution of scattering centers. Finally, we see that the large pathlength approximation breaks down for small pathlengths $\sim 1$ fm and even for large pathlengths $\sim5$ fm for large enough $\gtrsim100$ GeV energies.  We noted in \cref{sub:radiative_energy_loss} one more assumption, that of large transverse area.  This assumption is very difficult to assess using the methods of this article, especially as the utilization of the assumption occurs very early in the derivation.  Qualitatively, even in $\mathrm{p}+\mathrm{A}$ collisions, one has that the transverse size of the system will be $\sim(1\,\mathrm{fm})^2$ whereas the typical scale set by the scattering process itself is $1/\mu^2\sim(0.5\,\mathrm{fm})^2$.  It thus seems likely that this large transverse size assumption holds even for small collision systems.

Instead of thinking of the self-consistency of the numerics with the assumptions that went into the derivation of the energy loss, one may rather formulate the issue as whether or not one is integrating the matrix element (modulus squared) beyond the region under which its derivation is under control.  Thus one way of understanding that, e.g., $\langle \omega_0/\mu_1\rangle>1$ as shown in \cref{fig:largeformationtime3} is that the matrix element is integrated over regions of kinematics under which the derivation is not under control.  One may consider restricting the kinematics that are integrated over to those for which the derivation is under control.  We find, for example, that the expectation of the collinear approximation $\langle k^-/k^+\rangle$ is self-consistently less than 1, but note that the gluon kinematics are restricted in such a way as to enforce collinearity $|\mathbf{k}|^{\text{max}}=2x(1-x)E \Leftrightarrow k^- < k^+$.  As was shown in \cite{Horowitz:2009eb,Armesto:2011ht}, the DGLV inclusive gluon emission kernel is \emph{not} under good control near the kinematic bound $k^-\sim k^+$.  One may thus consider restricting the gluon kinematics such that the large formation time assumption is respected, for example, by taking $|\mathbf{k}|^{\text{max}} = \text{Min}(\sqrt{2xE\mu_1},2x(1-x)E) \Leftrightarrow \omega_0 < \mu_1$ and $k^- < k^+$.  Then presumably one would find that $\langle \omega_0/\mu_1\rangle\lesssim1$.  However, as was shown in \cite{Horowitz:2009eb}, there would then be a significant sensitivity in the energy loss model predictions to the exact kinematic bound chosen.

We thus conclude that in order to confidently compare leading hadron suppression predictions in $\mathrm{A}+\mathrm{A}$ collisions at $\gtrsim100$ GeV or in $\mathrm{p}+\mathrm{A}$ collisions at $\gtrsim 10$ GeV from an energy loss model based on an opacity expansion of the single inclusive gluon emission kernel, future work is needed to re-derive the opacity expansion single inclusive gluon emission kernel with both the large pathlength and the large formation time approximations relaxed.  Further work of numerically implementing finite pathlength effects in elastic energy loss \cite{Djordjevic:2006tw,Wicks:2008zz} will also play an important role in any quantitative comparison of an energy loss model and leading hadron suppression in small systems such as $\mathrm{p}+\mathrm{A}$ collisions.

\section*{Acknowledgments}

The authors thank Isobel Kolb\'e for valuable discussions and Chun Shen for supplying hydrodynamic temperature profiles. WAH thanks the National Research Foundation and the SA-CERN Collaboration for support.  CF thanks the SA-CERN Collaboration for their generous financial support during the course of this work.

\bibliography{small_system,manual}
\end{multicols}

\end{document}